\documentclass{article}

\usepackage{arxiv}

\usepackage[utf8]{inputenc} 
\usepackage[T1]{fontenc}    
\usepackage[colorlinks=true,
            citecolor=blue,
            linkcolor=blue,
            urlcolor=cyan]{hyperref}
\usepackage{url}            
\usepackage{booktabs}       
\usepackage{bm}
\usepackage{amsfonts}       
\usepackage{amsmath}
\usepackage{amssymb}
\usepackage{mathtools}
\usepackage{authblk}
\usepackage{romannum}
\usepackage{nicefrac}       
\usepackage{microtype}
\usepackage{cleveref}
\usepackage{xcolor}
\usepackage{algpseudocode}
\usepackage{algorithm}
\usepackage{color,soul}
\usepackage{caption}
\usepackage{lipsum}
\usepackage{fancyhdr}       
\usepackage{graphicx}       
\graphicspath{{media/}}     
\usepackage{enumitem}
\makeatletter
\newcommand{\printfnsymbol}[1]{%
  \textsuperscript{\@fnsymbol{#1}}%
}

\makeatother

\definecolor{elastomer}{HTML}{FF6B6B}
\definecolor{gel}{HTML}{90E0B0}

\usepackage[normalem]{ulem}

\definecolor{DLblue}{RGB}{0,0,180}

\usepackage{cite}

\title{GELATO: Multi-Material Topology Optimization of Programmable Gel-Elastomer Structures}

\author[+,1]{Aaditya Chandrasekhar}
\author[+,1]{Dex Doksoo Lee}
\author[1]{Hyunwoo Kwon}
\author[*,1]{Wei Chen}

\affil[1]{Department of Mechanical Engineering, Northwestern University, Evanston, IL, USA
\authorcr
   \{\tt aadityacs, weichen\}@northwestern.edu,
   \{\tt dslee, hyunwookwon2030\}@u.northwestern.edu}

\affil[+]{Equal contribution}
\affil[*]{Corresponding author}

\begin{document}
\maketitle

\begin{abstract}
Gel-elastomer composites, comprising an active swellable hydrogel and a passive elastomer, are a compelling class of programmable material systems (PMS) capable of shape morphing under multiphysics actuation. The precise design of the topology and material distribution unlocks complex programmability instrumental in wearable
electronics, soft robots, and drug delivery; however, the structure-function relationship is highly non-intuitive, rendering both trial-and-error and conventional design approaches largely intractable. To address this, we present a topology optimization (TO) framework for the automated design of such structures, enabling systematic exploration of the design space for target functionalities realized via programmable shape morphing. In particular, we propose a multi-material TO framework that concurrently optimizes the structural topology and the spatial distribution of the gel-elastomer phases. The design is represented via a coordinate-based neural network, and the mechanical response of both phases is described within a unified constitutive framework based on the Flory–Rehner theory. Furthermore, we present an end-to-end differentiable design framework with implicit differentiation that accommodates various objective functions, constraints, and discretizations. We demonstrate the framework on shape-programming structures and soft actuators. The framework is further validated through the design of organogel-hydrogel composites for multi-stimuli responsiveness across chemically distinct solvent environments, and of anisotropic hydrogels wherein the local fiber orientation is optimized concurrently with the topology. The codebase implemented in JAX is publicly shared to support benchmarking and reproducibility.

\end{abstract}


\section{Introduction}
\label{sec:intro}

Programmable material systems (PMS) form a class of material systems that can adapt their shapes or properties in response to external stimuli. Seen as key to embodied intelligence, this functional switch holds promise for applications that demand flexibility and robustness admist dynamic environments. Advances in manufacturing and fabrication and stimuli-responsive materials~\cite{truby2016PrintingSoft,wallin3DPrintingSoft2018,kuang2019Advances4da, yue2023single, zeng2023high, chen2025hybrid, oh2026architected, ze2020magnetic, ma2020magnetic} have fueled the recent development of PMS, supporting demonstrations in wearable electronics, haptic interfaces, soft robots, waveguiding, and others~\cite{wehner2016IntegratedDesign, rus2015DesignFabrication, kim2019FerromagneticSoft, ware2015VoxelatedLiquid,boley2019ShapeshiftingStructured,leanza_elephant_2024,luo2023autonomous, zhao2023encoding, wang2025direct, zhao2026dual}.

Among many kinds of PMS, gel-elastomers, a class of soft material composites, are increasingly utilized in stimuli-responsive systems in applications including soft robotic actuators, compliant mechanisms, microfluidic sensors, and drug delivery carriers \cite{puza2022hydrogelPrintingProgramming, boroomand2025opticalSensingHydrogel}. These composites combine two distinct phases: an active hydrogel that swells volumetrically in the presence of a solvent, and a passive elastomer that resists and redirects structural deformation. By optimally distributing these two phases, complex deformation patterns and functionalities can be programmed into the structure.

Despite their utility, manually designing hydrogel-elastomer structures for advanced applications is challenging owing to the complex structure-function relationships. Achieving desired programmability in soft material systems entails a challenging inverse problem, often involving joint consideration of geometry, material, stimulus, process, under dynamic environments. As such, the design of gel-elastomer systems presents challenges in coupled mechanics involving solvent diffusion and large hyperelastic deformations, complex boundary conditions (e.g., contact and friction), alongside mechanical and manufacturing constraints \cite{chester2015gelFEA}. This design is often further compounded by coupling among the design entities, multifunctionality requirements, multiphysics modeling, resource-intensive performance evaluation, and sensitivity analyses, particularly when gradient-based design is of interest.  

 \begin{figure}[H]
 	\begin{center}
		\includegraphics[scale=0.5,trim={0 0 0 0},clip]{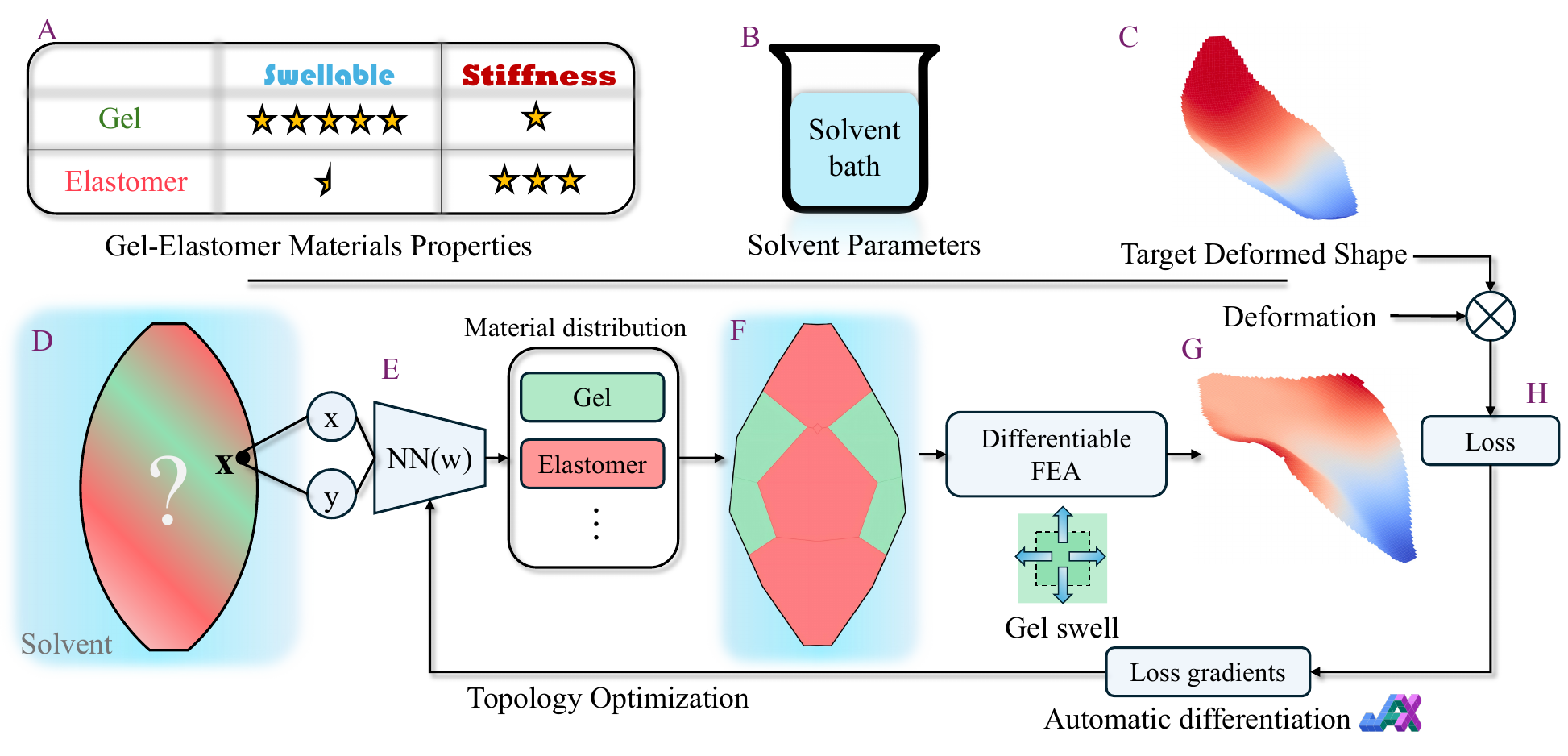}
 		\caption{Overview of the proposed framework, GELATO (\underline{G}el-\underline{Ela}stomer \underline{T}opology \underline{O}ptimization). Given (A) material properties of the two candidate phases, a swellable hydrogel and a stiff non-swellable elastomer, (B) solvent bath conditions that drive gel swelling, and (C) a target deformed shape, the framework optimizes the spatial distribution of the two materials within (D) a design domain that morphs upon solvent immersion as the gel phase selectively expands. A neural network (E) maps spatial coordinates to material composition at each point in the domain, producing (F) a multi-material layout whose swelling-induced deformation is predicted via (G) a differentiable FEA solver. The difference between the simulated and target shapes serves as the objective in (H) a topology optimization loop. The gradient of the loss are computed using automatic differentiation using JAX. The optimizer iteratively refines the material distribution until the design morphs to the target configuration upon swelling.}
 	\label{fig:fig_method_graphicalAbstract}
	\end{center}
 \end{figure}

To this end, we present GELATO (\underline{G}el-\underline{ELA}stomer \underline{T}opology \underline{O}ptimization), a multi-material topology optimization (TO) framework for the automated design of programmable gel-elastomer composites\cite{sigmund2013topoptreview}. An overview of the framework is illustrated in \Cref{fig:fig_method_graphicalAbstract}. We integrate three components to enable the concurrent optimization of the structural topology and the spatial distribution of the hydrogel and elastomer phases. First, we formulate an unified constitutive framework based on static, hyperelastic non-linear large deformation theory. Specifically, the Flory-Rehner theory \cite{chester2015gelFEA,hong2008GelTheory,flory1943FloryRhener} is employed to model the hydrogel swelling. Second, we introduce a coordinate-based neural network to map the spatial distribution of the material, allowing for a mesh-independent representation that naturally accommodates multi-material design constraints. Third, we construct an end-to-end differentiable simulation in JAX \cite{bradbury2018jax}, enabling the exploration of diverse objective functions and constraints through gradient-based optimization.

The remainder of this paper is organized as follows. \Cref{sec:litrev} reviews related work. \Cref{sec:method} details the governing equations, the material model, the domain description, and the optimization framework. Subsequently, \Cref{sec:expts} showcases the numerical experiments, demonstrating the utility of the framework on programmable shape morphing (\Cref{sec:expts_shapeMorphing}), relevant to soft robotic locomotion \cite{jiao2022programmableSoftRoboticsHydrogel} and 4D printing \cite{sydneygladman2016Biomimetic4d}; soft actuators (\Cref{sec:expts_inverter}), relevant to soft grippers \cite{puza2022hydrogelPrintingProgramming} and biomedical devices \cite{wehner2016IntegratedDesign}; organogel-hydrogel composites (\Cref{sec:expts_organohydrogels}), applicable to chemical sensing and drug delivery \cite{boroomand2025opticalSensingHydrogel}; and design with anisotropic hydrogels (\Cref{sec:expts_anisotropichydrogels}), relevant to bioinspired 4D printing \cite{sydneygladman2016Biomimetic4d}. Finally, \Cref{sec:conclusion} concludes the paper with future directions.

\section{Related Work}
\label{sec:litrev}


\textbf{Constitutive theory of gel swelling.} Flory and Rehner established the foundational theory of crosslinked gel swelling, deriving the first closed-form prediction from a free-energy formulation \cite{flory1943FloryRhener}. Subsequent work \cite{hong2008GelTheory,chester2015gelFEA} extended this thermodynamics into a coupled diffusion-deformation theory and transient finite-element implementation. The design goal for GELATO is to determine an optimal material configuration that achieves a target final deformation profile; we therefore adopt the equilibrium form, which collapses the chemo-mechanical equations to a static balance and avoids the cost of time-stepped integration.


\textbf{Inverse design of programmable material systems.} The inverse problem of mapping a material distribution to a target programmability has been addressed through analytical, parametric, data-driven approaches, and gradient-free optimization, each with constraints exposed by the gel-elastomer composite setting. For restricted geometry classes, closed-form relations invert this mapping analytically, expressing the required swelling, growth, or director field as a direct function of the target shape \cite{vanrees2017GrowthPatterns, aharoni2018universal}. Such methods are fast and physically interpretable but tractable only in idealized regimes that exclude the finite-strain kinematics and multi-physics coupling intrinsic to swelling gel-elastomer structures. Parametric designs represent a structure through a small set of tunable parameters, typically dimensions or material assignments, within a pre-specified topological class such as bilayer hydrogel composites~\cite{sydneygladman2016Biomimetic4d, stoychev2012shape}, multi-material rib lattices~\cite{boley2019ShapeshiftingStructured}, and voxel-patterned ferromagnetic or liquid-crystal composites~\cite{kim2019FerromagneticSoft, ware2015VoxelatedLiquid}. Data-driven approaches are receiving increasing attention in designing PMS as they relax analytical tractability and support on-the-fly predictions at inference~\cite{sun_machine_2024, lee_deep_2024, liwei2025autonomousStimuli}. However, the line presents grand challenges in data requirement with resource-intensive FE analysis, out-of-distribution generalization, and prediction accuracy associated with complex boundary conditions. Gradient-free optimization methods such as metaheuristic optimization and Bayesian optimization are applicable yet their adoption for PMS design is largely limited by scalability of high-dimensional design, convergence, and constraint handling. It remains elusive to exploit gradient-based design of hydrogel systems subject to the chemo-mechanical physics, which becomes increasingly relevant as inverse design problems grow in dimensionality and physical complexity.



\textbf{Differentiable simulation for gradient-based design.} Gradient-based TO of coupled chemo-mechanical physics requires sensitivities through a nonlinear forward solve. The re-derivation cost of the adjoint for each new constitutive law motivates an end-to-end differentiable simulation, in which gradients through the forward physics are computed by construction. This enables gradient-based optimization that scales to high-dimensional design spaces, converges in fewer iterations, and uses first-order stationarity as a stopping criterion. Examples include traditional adjoint-based methods, automatic differentiation in ML platforms (e.g., PyTorch and JAX), and differentiable simulators (e.g., NVIDIA Warp~\cite{macklin2022warp}, DiffTaichi~\cite{hu2019difftaichi},  GradSim~\cite{jatavallabhula2021gradsim}, Dojo~\cite{howell2022dojo}, PhiFlow~\cite{holl2020phiflow}). Those are increasingly adopted by gradient-based design of PMS~\cite{bordiga2024automated, wang2025CodesignMagnetic, yuhn20234d}.

\textbf{Design representation in TO.} Within gradient-based TO, the design representation determines what topologies the optimizer can express and how multi-material constraints are encoded. Density-based methods (SIMP) \cite{bendsoeSIMP, sigmund2013topoptreview} take element-wise pseudo-densities as design variables, with effective material properties obtained via power-law interpolation in those densities. Level-set methods \cite{wang2003levelset, allaire2004levelsetsensi} represent material boundaries implicitly through a signed-distance function whose zero contour defines the design. Moving morphable components \cite{Guo2014MMC} parameterize the design through a fixed set of geometric primitives. These formulations have been broadly applied to single-physics structural TO. Neural network representations have recently emerged as an alternative, parameterizing the design field through a neural network of spatial coordinates~\cite{zehnder2021ntopo, hoyer2019neural,nobari2024nito}. These representations yield mesh-independent design fields and compact parameterizations. We build on the TOuNN family~\cite{chandrasekhar2021tounn,chandrasekhar2021multi,chandrasekhar2022approximate} and extend it to joint topology and three-phase material parameterization for swelling gel-elastomer composites.


\section{Proposed Method}
\label{sec:method}

 \begin{figure}[]
 	\begin{center}
		\includegraphics[scale=0.55,trim={0 0 0 0},clip]{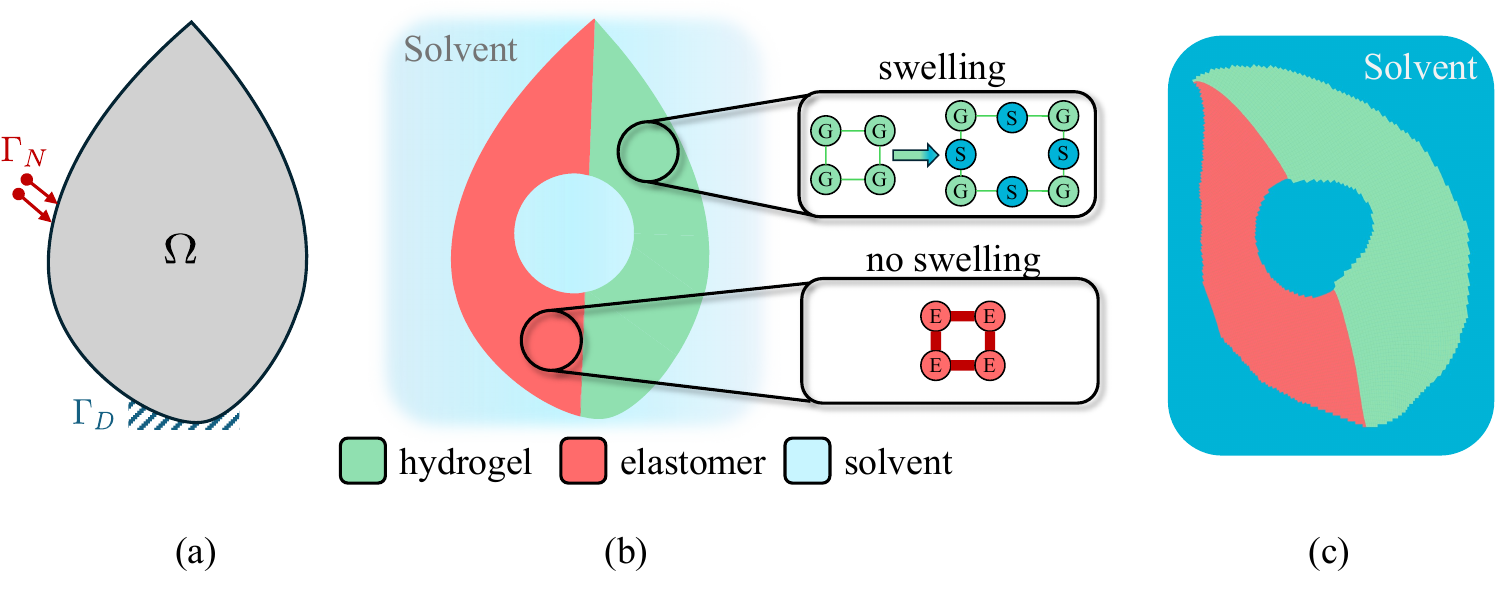}
 		\caption{Problem setup. (a) Design domain $\Omega$ with Dirichlet and Neumann boundary conditions. (b) The hydrogel phase swells upon solvent exposure while the elastomer phase resists swelling; the void phase is occupied by the solvent. (c) Deformation of the gel-elastomer composite.}
 		\label{fig:fig_method_problem_setup}
	\end{center}
 \end{figure}

We propose a topology optimization (TO) framework for the design of hydrogel-elastomer composites, simultaneously optimizing topology and material distribution. The structure combines an active, swellable hydrogel with a passive elastomer and is actuated by solvent immersion.

In particular, the hydrogel phase is assumed to be mechanically soft and to possess thermodynamic affinity to the solvent. This drives solvent diffusion into the gel, as dictated by the gradient in chemical potential between the surrounding medium and the reference dry state of the hydrogel, producing large mechanical deformations. In contrast, the elastomer phase is mechanically stiffer and possesses no thermodynamic affinity to the solvent, causing no swelling upon immersion \cite{chester2015gelFEA}. Instead, it functions as a structural stiffener, resisting and redirecting the swelling-induced deformations of the hydrogel. In general,
composing a structure from these two phases enable spatially non-uniform deformation profiles determined entirely by the material distribution. By employing an optimizer to precisely determine the topology and spatial placement of the hydrogel and elastomer, the proposed framework enables the programming of complex deformation profiles, realizing structures with targeted shape-morphing and prescribed mechanical functions (\Cref{fig:fig_method_problem_setup}).

The optimization is formulated over a prescribed design domain with associated boundary conditions, objective, and design constraints. Additionally, the framework takes as input the relevant structural properties (shear modulus), environmental conditions (temperature and chemical potentials), fluidic properties (solvent molar volume), and swelling-specific parameters (the Flory–Huggins interaction parameter) for both constituent phases. A density-based multi-material TO scheme \cite{sanders2018MMTO} is then employed to concurrently determine the topology and the spatial assignment of the hydrogel and elastomer throughout the domain.

The remainder of this section is organized as follows. The governing physics are detailed in \Cref{sec:method_governingEquations}. The neural network based design representation and material interpolation schemes are described in \Cref{sec:method_materialModel}. The finite element schemes employed are briefly discussed in \Cref{sec:method_fem}. The complete optimization formulation, loss function, sensitivity analysis, and the optimization procedure are presented in \Cref{sec:method_optimization} and \Cref{sec:method_sensAnalysis}. Finally, the complete algorithm is summarized in \Cref{sec:method-algo}.

\subsection{Governing Equations}
\label{sec:method_governingEquations}

This section presents a brief overview of the theoretical model adopted to describe the mechanics of the hydrogel-elastomer structure. In particular, we adopt the theoretical framework of \cite{hong2008GelTheory}, which integrates large-deformation mechanics with the thermodynamics of solvent-hydrogel interactions. The analysis is restricted to the fully equilibrated static response. We assume the gel to have exchanged solvent with the surrounding medium until the chemical potential is homogeneous and equal to that prescribed by the external environment. Furthermore, we assume molecular incompressibility, while the bulk material remains macroscopically compressible. Finally, we assume isothermal conditions, where the solvent bath maintains a constant temperature throughout the swelling process.

We adopt a unified constitutive model for both the hydrogel and the elastomer. In practice, this corresponds to modeling the elastomer as a stiffer hydrogel with negligible affinity for the solvent. This unified polymeric description allows continuous interpolation between the elastomer and hydrogel, paving the way for gradient-based design optimization.

We begin the analysis by considering the design domain $\Omega$ in its reference configuration. This domain is eventually partitioned, through TO, into non-overlapping elastomer ($\Omega_E$), hydrogel ($\Omega_G$), and void ($\Omega_V$) subdomains, such that their union forms the complete design domain: $\Omega = \Omega_E \cup \Omega_G \cup \Omega_V$. Furthermore, we assume prescribed Dirichlet boundary conditions on $\Gamma_D$ and Neumann boundary conditions on $\Gamma_N$, such that $\Gamma_D \cup \Gamma_N = \partial\Omega$ and $\Gamma_D \cap \Gamma_N = \emptyset$.

The design undergoes deformation such that points in the reference configuration $\bm{X} \in \Omega \subset \mathbb{R}^d$ are mapped to the current configuration $\Omega'\subset \mathbb{R}^d$ via a deformation map $\zeta: \bm{X} \rightarrow \bm{x}$ \cite{neto2008computationalMethods}. The associated deformation gradient tensor is defined as $\bm{F} = \nabla_0 \zeta$, where $\nabla_0$ denotes the gradient operator with respect to the reference configuration. We require $J = \det \bm{F} > 0$ to ensure a physically valid, orientation-preserving mapping. We adopt the multiplicative decomposition $\bm{F} = \bm{F}^e \bm{F}^s$, where $\bm{F}^s$ is the swelling part and $\bm{F}^e$ is the elastic part. Furthermore, we track the amount of solvent uptake by defining a polymer volume fraction $\phi(\bm{x}) \in (0,1]$, where $\phi = 1$ corresponds to a completely dry polymer. Then with isotropic swelling, we have $\bm{F}^s = \phi^{-1/3}\bm{1}$, where $\bm{1}$ is the identity tensor.  Furthermore, molecular incompressibility enforces $J^e = \det\bm{F}^e = 1$, so that the total volumetric Jacobian $J = \det\bm{F}$ is entirely governed by solvent uptake, yielding $J = J^eJ^s = \phi^{-1}$.

The elastic strain energy density of the gel takes the Neo-Hookean form \cite{hong2008GelTheory}:
\begin{equation}
  \psi = \frac{G}{2}\left(I_1 - 3 - 2\ln J\right),
  \label{eq:free_energy}
\end{equation}
where $G$ is the shear modulus and $I_1$ is the first principal invariant of the right Cauchy-Green tensor, and the $-2\ln J$ term accounting for macroscopic volume change due to solvent absorption. Under our plane stress assumption, $I_1$ can be expressed in terms of the in-plane principal stretches ($\lambda_1$ and $\lambda_2$) and the polymer volume fraction:
\begin{equation}
  I_1 = \lambda_1^2 + \lambda_2^2 + \frac{1}{\phi^2 \lambda_1^2 \lambda_2^2}.
  \label{eq:I1}
\end{equation}

The polymer volume fraction $\phi$ is determined (implicitly via a bisection solver) by the thermodynamic equilibrium condition between the gel and the external solvent, as dictated by the nonlinear Flory-Rehner equation \cite{flory1943FloryRhener}:
\begin{equation}
  \mathcal{F}(\phi) \coloneqq \frac{\mu^0 - \mu}{RT} + \ln(1-\phi) + \phi + \chi\phi^2
  + \frac{\tilde{\Omega} G}{RT}\
  \left(\frac{1}{\phi\lambda_1^2\lambda_2^2} -\phi\right) = 0,
  \label{eq:flory_rehner}
\end{equation}
where $\mu$ is the chemical potential of the external solvent, $\mu^0$ is its reference value, $R$ is the universal gas constant, $T$ is the absolute temperature, $\tilde{\Omega}$ is the molar volume of the solvent, and $\chi$ is the Flory-Huggins interaction parameter governing the affinity between the solvent and the material. Intuitively, the first term in \Cref{eq:flory_rehner} dictates the chemical driving potential, the next three terms represent the free energy of mixing, and the final term captures the elastic resistance of the swollen network to further solvent uptake.

 \begin{figure}[H]
 	\begin{center}
		\includegraphics[scale=0.45,trim={0 0 0 0},clip]{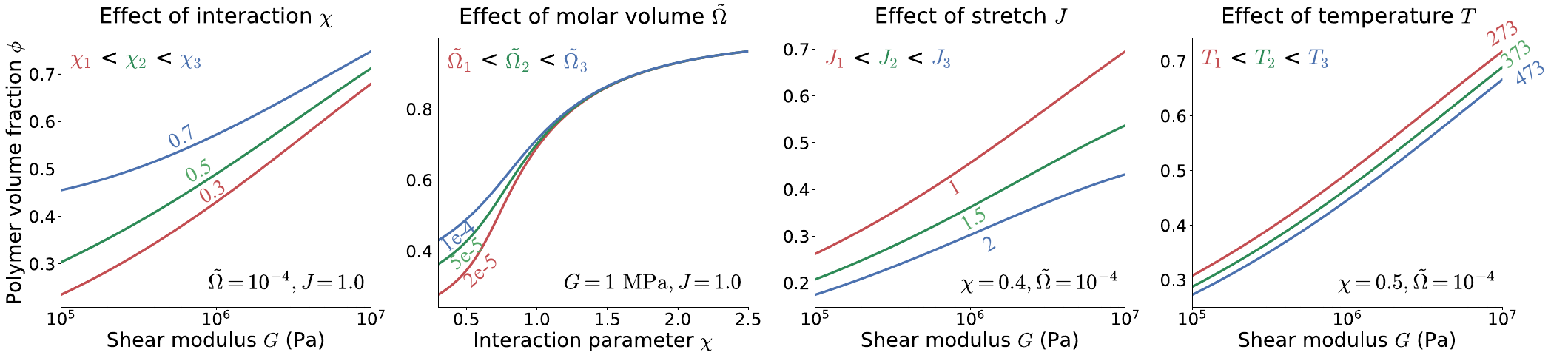}
 		\caption{Effect of the governing parameters on the polymer volume fraction.}
 		\label{fig:flory_rehner_param_effect}
	\end{center}
 \end{figure}

\Cref{fig:flory_rehner_param_effect} illustrates the dependence of $\phi$ on the governing parameters. Two trends are noteworthy. Firstly, observe that the interaction parameter $\chi$ exerts a strong influence on swelling: increasing $\chi$ reduces the thermodynamic affinity between the solvent and the polymer, markedly suppressing solvent absorption and driving $\phi \rightarrow 1$. This observation underpins a key modeling choice in this work: both void and elastomer phases are represented as hydrogels with a sufficiently large $\chi$, such that $\phi \approx 1$ and swelling is effectively absent, paving the way for a unified treatment. Secondly, the figure confirms that hydrogels, characterized by low shear moduli ($G \sim 10^4$--$10^5\,\text{Pa}$), exhibit substantially larger swelling than elastomers ($G \sim 10^6$--$10^7\,\text{Pa}$), consistent with the physical expectation that stiffer networks resist solvent absorption more strongly.

Having defined the constitutive model, the elastostatic equilibrium of the structure in the reference configuration is governed by:

\begin{equation}
  \begin{cases}
    -\nabla_0 \cdot \bm{P} = \bm{0}, & \text{in } \Omega, \\
    \bm{u} = \bar{\bm{u}},            & \text{on } \Gamma_D, \\
    \bm{P}\,\bm{n} = \bar{\bm{t}}, & \text{on } \Gamma_N,
  \end{cases}
  \label{eq:strong_form}
\end{equation}

where $\bm{P} = \partial\psi/\partial\bm{F}$ is the first Piola-Kirchhoff stress tensor, $\bar{\bm{u}}$ is the prescribed displacement on $\Gamma_D$, $\bar{\bm{t}}$ is the prescribed traction on $\Gamma_N$, and $\bm{n}$ is the outward unit normal in the reference configuration. Here, we neglect inertial effects and body forces. The corresponding weak form of the equilibrium, derived from the principle of minimum potential energy, seeks a kinematically admissible displacement field $\bm{u} \in \mathcal{U} \subset H^1(\Omega)$ such that:
\begin{equation}
  \int_{\Omega} \bm{P} : \nabla_0(\delta\bm{u})\,dV
  - \int_{\Gamma_N} \bar{\bm{t}} \cdot \delta\bm{u}\,dA = 0
  \label{eq:weak_form}
\end{equation}
for all virtual displacements $\delta\bm{u} \in \mathcal{V}_0 =
\{\bm{v} \in H^1(\Omega) : \bm{v}|_{\Gamma_D} = \bm{0}\}$. This variational equation is subsequently discretized utilizing standard finite element procedures. Both shell-based and quadrilateral-based element formulations are considered, as detailed in \Cref{sec:method_fem}.

\subsection{Material Model and Design Representation}
\label{sec:method_materialModel}

Having established the governing equations, we now outline the material model used in the optimization. Recall that the goal is to seek a disjoint partition of $\Omega$ into elastomer ($\Omega_E$), hydrogel ($\Omega_G$), and void ($\Omega_V$) subdomains. Consequently, we define the composition of the material at every point $\bm{x} \in \Omega$ through a set of pseudodensities of gel, elastomer, and voids $\{\rho_g, \rho_e, \rho_v\}$ respectively. For a physically realizable design, we require a disjoint partition that satisfies $\rho_{g,e,v}(\bm{x}) \in \{0,1\}\ \forall\, \bm{x}$ and $\rho_g + \rho_e + \rho_v = 1$. However, for continuous gradient-based optimization, we relax and allow $0 \leq \rho_g, \rho_e, \rho_v \leq 1$, driving the densities toward $\{0,1\}$ through penalization and additional constraints.

The effective shear modulus $G$ and Flory-Huggins parameter $\chi$ are then obtained via the standard Solid Isotropic Material with Penalization (SIMP) scheme \cite{sigmund2013TopologyOptimization}:
\begin{equation}
  G(\bm{x}) = \sum_{i\in\{g,e,v\}} \rho_i^p(\bm{x})\, G_i,
  \label{eq:eff_G}
\end{equation}
\begin{equation}
  \chi(\bm{x}) = \sum_{i\in\{g,e,v\}} \rho_i^q(\bm{x})\, \chi_i,
  \label{eq:eff_chi}
\end{equation}
where $p(=3)$ and $q(=1)$ are the penalization exponents, and $G_g$, $G_e$, $G_v$ and $\chi_g$, $\chi_e$, $\chi_v$ are the shear moduli and Flory-Huggins parameters of the hydrogel, elastomer, and void phases, respectively. Note that the higher penalization of $G$ suppresses intermediate modulus values, pushing the density field toward crisp phase boundaries, while $\chi$ is left linear ($q = 1$) since the nonlinearity of the Flory-Rehner equation already provides sufficient contrast in swelling response between phases. In practice, hydrogels are soft, while elastomers are stiffer. Following standard procedures in TO, the void phase is assigned a small residual modulus $G_v \approx 10^{-2}\,G_g$ to avoid numerical singularities while remaining negligible relative to the solid phases. The hydrogel interaction parameter is set at commonly used value $\chi_g \in [0.1, 0.5]$, reflecting strong affinity for the solvent and substantial swelling. The elastomer and void phases are assigned large values $\chi_e, \chi_v \gg 1$ to reflect inertness to the solvent medium.

In this work, we parameterize the design via a coordinate-based neural network \cite{chandrasekhar2021tounn,chandrasekhar2021multi}. In contrast to conventional TO, where design variables are collocated with the finite element discretization, this neural representation offers several  advantages for multi-material design: (a) the partition of unity requirement is intrinsically satisfied through the choice of output activation, requiring no additional enforcement; (b) the design field is decoupled from the simulation mesh, yielding a compact set of design variables that is independent of mesh resolution; (c) the material field is defined analytically throughout the domain, enabling the recovery of crisp, high-resolution designs in post-processing, and (d) spatial gradients $\nabla_{\bm{x}} \rho$ are computed exactly via backpropagation through the network, which is essential for resolving sharp material interfaces ; (e) the global parameterization inherently regularizes the design, preventing checkerboarding without requiring explicit spatial filters \cite{sigmund1998NumericalInstabilities}.

 \begin{figure}[]
 	\begin{center}
		\includegraphics[scale=0.6,trim={0 0 0 0},clip]{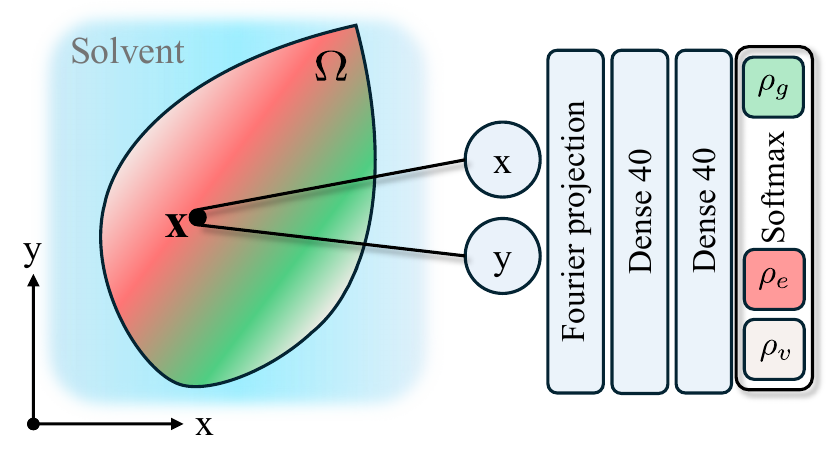}
 		\caption{The neural network maps spatial coordinates to the material distribution.}
    \label{fig:fig_method_nn_representation}
	\end{center}
 \end{figure}

The network is structured as follows. The input layer ingests the spatial coordinates $\bm{x} \in \mathbb{R}^d$. These are then passed through a Fourier projection layer \cite{chandrasekhar2022approximate,tancik2020fourier}, which lifts the Euclidean coordinates into a higher-dimensional Fourier space. This projection serves three purposes: (a) alleviates spectral bias of standard multilayer perceptrons toward low-frequency solutions \cite{rahaman2019spectral}, (b) enables the network with the capacity to represent high-frequency spatial variations in the pseudodensities, and (c) accelerates convergence of the optimization. The projected features are then processed through two dense hidden layers, each comprising 40 neurons with ReLU activations. The output layer consists of 3 neurons, one for each phase: hydrogel ($\rho_g$), elastomer ($\rho_e$), and void ($\rho_v$), with Softmax activation applied to the final layer. This ensures that $\sum\limits_{i \in \{g,e,v\}} \rho_i = 1$ and $0 \leq \rho_g, \rho_e, \rho_v \leq 1$ are satisfied by construction. The trainable weights $\bm{w}$ of this network constitute the design variables for optimization.

\subsection{Finite Element Analysis}
\label{sec:method_fem}

Having described the material model and design representation, we now detail the finite element analysis used to evaluate the structural response. The weak form \Cref{eq:weak_form} is discretized using two element formulations, depending on the problem setting. For planar problems, we employ quadrilateral bilinear elements. To mitigate volumetric locking arising from the near-incompressible swollen response, we adopt the $\bar{\bm{F}}$ method \cite{neto2008computationalMethods}. For thin shell problems, we employ the Naghdi shell formulation discretized with MITC-4 elements \cite{bathe1984shellmitc}, which mitigates shear locking. Each node is attributed with three translational and two rotational degrees of freedom, with the drilling rotation constrained. In general, we express the discrete residual equation governing equilibrium as:

\begin{equation}
    \bm{R}(\bm{u}) \coloneqq \bm{f}_{\mathrm{int}}\bigl(
    \bm{u},\, \phi(\bm{u})\bigr) - \bm{f}_{\mathrm{ext}} 
    = \bm{0},
    \label{eq:residual}
\end{equation}

where $\bm{u}$ denotes the vector of nodal degrees of freedom (displacements, and additionally rotations for shell elements), $\bm{f}_{\text{int}}$ is the internal force vector, and $\bm{f}_{\text{ext}}$ is the external force vector. Observe that the system is intrinsically coupled: $\phi$ depends on the current deformation state $\bm{u}$ through the in-plane stretches appearing in the Flory-Rehner \Cref{eq:flory_rehner}, which is solved via bisection at each Newton iteration. The displacement field $\bm{u}$ and the polymer volume fraction $\phi$ are resolved concurrently within each iteration, though only $\bm{u}$ is the Newton unknown.

Owing to the geometric nonlinearities and large deformations induced during swelling, the external stimuli are applied incrementally. In particular, the chemical potential of the solvent bath is stepped from the dry reference state to the prescribed environmental value $\mu$. Additionally, any external mechanical loads and non-homogeneous Dirichlet conditions are stepped concurrently over the same increments. At each load step, \Cref{eq:residual} is solved by a damped Newton-Raphson iteration with Armijo backtracking line search \cite{nocedal2006numericalOptimization}. Each load step is warm-started from the previously converged solution. Given our end-to-end differentiable framework, the consistent tangent stiffness matrix $\bm{K} = \partial\bm{R}/\partial\bm{u}$ is obtained via automatic differentiation of the residual through the full coupled system, including the dependence of $\phi$ on $\bm{u}$. The solver is implemented within the JAX ecosystem \cite{bradbury2018jax}, with PETSc \cite{dalcin2011petsc} employed as the linear solver at each Newton step.

\subsection{Optimization}
\label{sec:method_optimization}

Having established the governing equations, material models, design representation, and the analysis procedure, we now outline the optimization framework. The numerical experiments in \Cref{sec:expts} consider a variety of design objectives and constraints; we present the complete set here and invoke the relevant subset in each study.

\subsubsection{Design Variables}
\label{sec:method_optimization_designVars}

Recall that the topology and material distribution are parameterized by the weights $\bm{w}$ of the coordinate-based neural network. These weights thus constitute the design variables of our optimization. At each iteration, the coordinates of the mesh element centers or the integration points are passed through the network to evaluate the topology and material pseudodensities $ \bm{\rho}(\bm{x}) = \mathcal{N}(\bm{x};\, \bm{w})$. Recall that the Softmax output activation intrinsically enforces the partition of unity constraint $\rho_g + \rho_e + \rho_v = 1$ and the bounds $0 \leq \rho_{g,e,v} \leq 1$. The weights are initialized using the Xavier normal scheme, corresponding to a random initial pseudodensity field.

\subsubsection{Objective \Romannum{1}: Shape Morphing}
\label{sec:method_optimization_obj_shapeMorphing}

One of the central goals of the case studies showcased herein (\Cref{sec:expts_shapeMorphing,sec:expts_anisotropichydrogels}) is computing the optimal material distribution such that the structure morphs to a target configuration upon swelling. Given a target displacement field $\bm{u}^{tgt}(\bm{x})$ and the FEA simulated displacement field $\bm{u}(\bm{x};\,\boldsymbol{\rho})$ at the current design iterate, the objective is expressed as the minimization of the mean squared error (MSE) between the two fields, evaluated at a set of $n_s$ target points $\{\bm{x}_k\}_{k=1}^{n_s}$:
\begin{equation}
  J(\boldsymbol{\rho}) 
    = \frac{1}{n_s}\sum_{k=1}^{n_s} 
      \left\|\bm{u}(\bm{x}_k;\,\boldsymbol{\rho}) - \bm{u}^{tgt}(\bm{x}_k)\right\|^2.
  \label{eq:objective_morphing}
\end{equation}

\subsubsection{Objective \Romannum{2}: Mechanisms}
\label{sec:method_optimization_obj_mechanisms}

Another goal of the case studies showcased (\Cref{sec:expts_inverter,sec:expts_organohydrogels}) is the design of swelling-actuated compliant mechanisms, where the structure is optimized to generate a directed mechanical output at a prescribed output boundary $\Gamma_\mathrm{out}$. Towards this, we adopt the blocked force as the design metric, which is the reaction force at $\Gamma_\mathrm{out}$ when the output port is held at zero displacement. The blocked condition is enforced by prescribing zero displacement Dirichlet condition at the output degrees of freedom. The objective is then to maximize  the net reaction force at $\Gamma_\mathrm{out}$ at mechanical equilibrium ($J_\mathrm{BF}$):
\begin{equation}
  J_\mathrm{BF}(\bm{\rho}) 
    = \bm{l}^\top \bm{f}_\mathrm{int}(\bm{u}^*(\bm{\rho})),
  \label{eq:objective_blocked_force}
\end{equation}
where $\bm{l} \in \{0, \pm 1\}^{n_\mathrm{dof}}$ is an indicator vector selecting the output degrees of freedom in the prescribed actuation direction, and $\bm{f}_\mathrm{int}$ is the internal force vector at the converged displacement $\bm{u}^*$.

\subsubsection{Volume Constraint}
\label{sec:method_optimization_volCons}

In some design scenarios, a volume constraint is imposed on a phase or a subset of phases $\mathcal{P} \subseteq \{g, p, s\}$, where $s$ denotes the aggregate solid phase with $\rho_s = \rho_g + \rho_e$. For compliant mechanism design, this is necessary to avoid trivial solutions. For lightweight structure design, it serves to limit material usage. Let $v_e$ denote the volume of element $e$ and $\rho_m(\bm{x}_e)$ the pseudodensity of phase $m$ at element center $\bm{x}_e$. For each $m \in \mathcal{P}$, the constraint is expressed as:
\begin{equation}
  g_v^m \coloneqq \frac{\sum\limits_e \rho_m(\bm{x}_e)\,v_e}{\sum\limits_e v_e} 
  \leq V_m^*,
  \label{eq:vol_constraint}
\end{equation}
where $V_m^* \in (0,1]$ is the prescribed upper bound on the volume fraction 
of phase $m$.

\subsubsection{Grayness Constraint}
\label{sec:method_optimization_grayCons}

Recall (\Cref{sec:method_materialModel}) that while the pseudodensity field was relaxed to a continuous domain ($0 \le \rho_{g,e,v}(\bm{x}) \le 1$) to enable gradient-based optimization, physical realizability dictates that the final layout must be strictly discrete ($\rho_{g,e,v}(\bm{x}) \in \{0, 1\}$). To drive the design toward this binary state and eliminate intermediate pseudodensities, we introduce a grayness constraint \cite{qian2017undercut}:

\begin{equation}
g_{r} \coloneqq \frac{1}{n_e} \sum\limits_{i}^{n_e} \sum\limits_{k \in \{g,e,v\}}\rho_{k}(\bm{x}_i) (1 - \rho_{k}(\bm{x}_i)) \le  \xi ,
\label{eq:grayness_constraint}
\end{equation}

where $\bm{x}_i$ ($i=1,\ldots,{n_e}$) are the sampled spatial coordinates, often collocated with the element centroids of the finite element discretization. Furthermore, $\xi > 0$ serves as a relaxation bound governed by a continuation scheme. Initially, $\xi$ is assigned a sufficiently large value to permit mixed material phases, enabling the optimizer to broadly explore the design space without prematurely trapping in local minima. Over the course of the optimization, $\xi$ is systematically annealed toward zero. This progressive tightening imposes a strict penalty on gray regions, effectively resolving the design to a discrete partition.

\subsubsection{Optimization Formulation}
\label{sec:method_optimization_loss}

Collecting the objective, residual (\Cref{eq:residual}), constraints the canonical optimization problem can be expressed as:

\begin{subequations}
	\label{eq:optimization_nn_Eqn}
\begin{align}
    \underset{\bm{w}}{\text{minimize}} \quad & J \\
    \text{subject to} \quad & \bm{R}(\bm{u}) = 0 \\
    & g_i \le 0 \; i \in \{r,v \} \\
\end{align}
\end{subequations}

The constrained minimization problem in \Cref{eq:optimization_nn_Eqn} is then transformed into an unconstrained loss function minimization, using the log-barrier scheme \cite{kervadec2022constrained}. Specifically, the loss function is defined as

\begin{equation}
\mathcal{L} = \frac{J}{J_0} + \sum_i\psi(g_i),
\label{eq:loss}
\end{equation}

where,

\begin{equation}
   \psi_\tau(g) = \begin{cases}
-\frac{1}{\tau} \ln(-g), & g \leq -\frac{1}{\tau^2} \\
\tau g - \frac{1}{\tau} \ln(\frac{1}{\tau^2}) + \frac{1}{\tau}, & \text{otherwise}
\end{cases} 
\end{equation}

and $J^0$ is the initial objective.
The constraint penalty parameter $\tau$ is updated at each optimization iteration $'k'$ as $\tau = \tau_0 \nu^k$ (where, $\tau_0 = 3$ and $\nu = 1.03$), making the enforcement of the constraint stricter as the optimization progresses.
\subsection{Sensitivity Analysis}
\label{sec:method_sensAnalysis}

Following the optimization formulation, we detail the computation of the sensitivities of the loss, objective, and constraints with respect to the design variables. Computing these sensitivities is non-trivial in the present setting, owing to geometric nonlinearity, the intrinsic coupling between deformation and swelling, and the integration of a neural network design representation with a nonlinear FEA solver. To address this, we construct an end-to-end differentiable pipeline using the reverse-mode automatic differentiation (AD) capabilities of JAX \cite{bradbury2018jax}. This obviates the laborious and error-prone process of manually deriving sensitivity expressions and computes all required derivatives to machine precision. Furthermore, the framework naturally accommodates varying discretization schemes, objectives, and constraints in a plug-and-play manner.

Two challenges specific to our simulation warrant further discussion. The first concerns the iterative root finding solvers employed: the Newton-Raphson method for the displacement field and bisection for the Flory-Rehner equation. A naive application of AD would unroll derivative computation across every iteration, incurring prohibitive memory and compute costs. We instead apply the implicit function theorem \cite{blondel2022implicitdiff}, which computes derivatives directly from the converged solution without backpropagating through the iteration history. We refer the readers to the Appendix for a detailed treatment on this. The second challenge arises from load stepping. An adjoint sensitivity computation requires access to the deformation state at every load step, and storing this full history is memory-intensive for fine discretizations with many increments. We address this via a checkpointing scheme \cite{wang2009checkpointing}, wherein the state is stored only at selected load steps and intermediate states are recomputed on the fly during the adjoint pass. This significantly reduces memory requirements at a moderate increase in computation time.

\subsection{Algorithm}
\label{sec:method-algo}

Having defined all components of the framework, we now summarize and present the complete algorithm.

\begin{enumerate}

\item{\textbf{Domain Initialization:}}
The computational domain $\Omega$ and the FEA mesh are defined. Dirichlet boundary conditions $\bar{\bm{u}}$ are prescribed on $\Gamma_D$ and Neumann boundary conditions $\bar{\bm{t}}$ are prescribed on $\Gamma_N$, as described in (\Cref{sec:method_governingEquations}).

\item{\textbf{Material Initialization:}}
The material constants are specified. These include the shear moduli $G_g$, $G_e$, and $G_v$ and Flory-Huggins interaction parameters $\chi_g$, $\chi_e$, and $\chi_v$ of the hydrogel, elastomer, and void phases, respectively. Furthermore, the environmental parameters including the solvent molar volume $\tilde{\Omega}$, the absolute temperature $T$, the reference chemical potential $\mu^0$, and the chemical potentials of the dry and fully swollen solvent baths $\mu_{\mathrm{dry}}$ and $\mu_{\mathrm{wet}}$ are specified.

\item{\textbf{Network Initialization:}}
The coordinate-based network $\mathcal{N}(\cdot\,;\bm{w})$ is instantiated. The weights $\bm{w}$ are initialized using the Xavier normal scheme, yielding a random initial pseudodensity field. The bandwidth of the Fourier projection layer, which controls the feature sizes of the material distribution, is also defined.

\item{\textbf{Solvers and optimizer:}}
The tolerances and maximum iteration counts for the bisection solver, Newton-Raphson iteration, and linear solver are specified. The optimization settings are then prescribed including the log-barrier penalty parameter, the SIMP penalization exponents $p, q$, the threshold filter parameters, the grayness constraint slackness, and the Adam optimizer hyperparameters (learning rate and gradient clip norm). Note that we adopt continuation schemes for the SIMP penalization, threshold filter, and the grayness constraint in our work (\cite{sigmund1998NumericalInstabilities}).

\item{\textbf{Material pseudo-densities:}}
At each iteration, the coordinates of the element centers and/or integration points $\{\bm{x}\}$ are passed through the coordinate-based network to generate the pseudo-density field: $\bm{\rho}(\bm{x}) = \mathcal{N}(\bm{x};\, \bm{w}),$ yielding the local volume fractions $\rho_g$, $\rho_e$, and $\rho_v$ at each point, as described in \Cref{sec:method_materialModel}.

\item{\textbf{Property mapping:}}
The pseudodensities are used to compute the spatially varying effective material properties via \Cref{eq:eff_G,eq:eff_chi}, mapping the density field to the effective shear modulus $G(\bm{x})$ and Flory-Huggins parameter $\chi(\bm{x})$ at each queried point.

\item{\textbf{Finite element analysis:}}
The nonlinear static equilibrium equations to determine the swelling induced deformations is solved using the procedure detailed in \Cref{sec:method_fem} and summarized in Algorithm 1. Given the nonlinearities and large deformation, we adopt a load-stepping protocol where the external stimuli are applied incrementally. In particular, the chemical potential is stepped from $\mu_{\mathrm{dry}}$ to $\mu_{\mathrm{wet}}$, with any external structural loads and non-homogeneous Dirichlet conditions stepped concurrently over the same increments. Furthermore, both bilinear quadrilateral and shell-based element formulations are utilized in our studies. The analysis yields the nodal displacement field $\bm{u}$.

\item{\textbf{Loss and constraints:}}
The relevant objective $J$ and constraints $\{g\}$ are evaluated using the pseudodensities and displacement field $\bm{u}$. The total loss $\mathcal{L}$ is computed by aggregating the objective and constraints via the log-barrier scheme, as described in \Cref{sec:method_optimization_loss}.

\item{\textbf{Sensitivity analysis:}}
The gradient $\partial \mathcal{L} / \partial \bm{w}$ is computed via reverse-mode automatic differentiation through our end-to-end differentiable pipeline, as described in \Cref{sec:method_sensAnalysis}. Utilizing the computed sensitivities, the design variables $\bm{w}$ are updated using the Adam optimizer. The parameters incorporating a continuation scheme are then advanced. 

\item{\textbf{Update:}}
The steps 5 through 9 are repeated until the convergence criterion is satisfied or the maximum number of iterations is reached.
\end{enumerate}

\begin{figure}[H]
 	\begin{center}
		\includegraphics[scale=0.55,trim={0 0 0 0},clip]{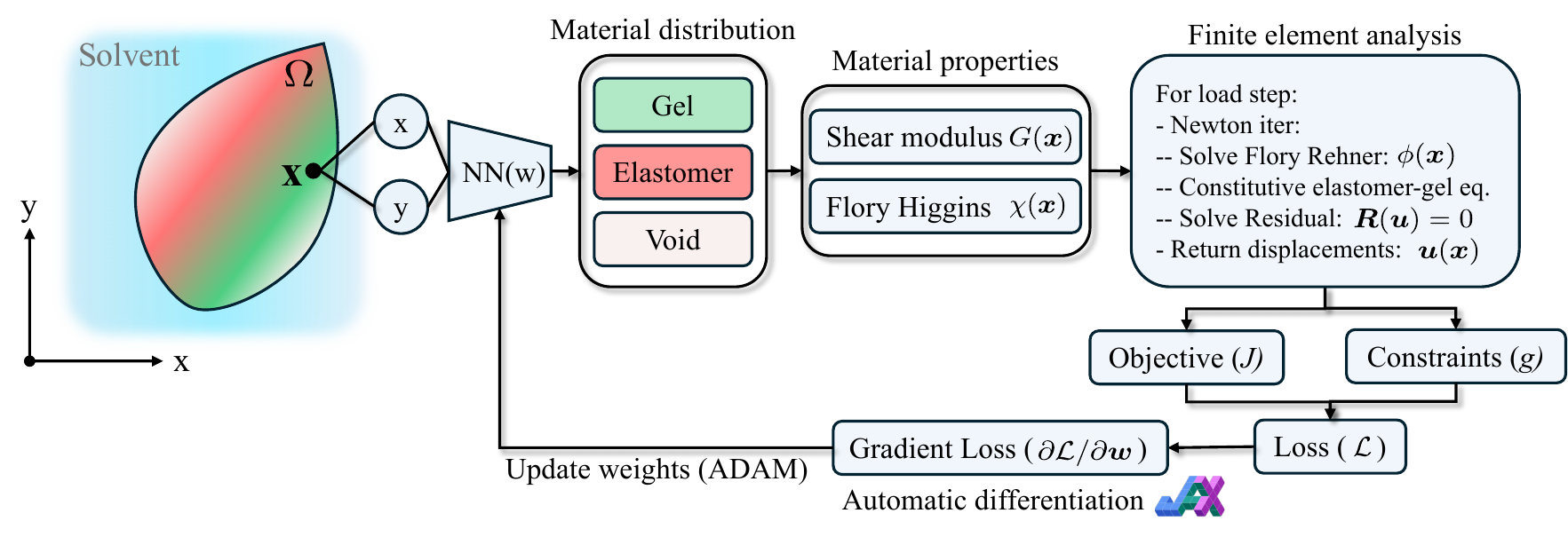}
 		\caption{Overview of the optimization loop.}
    \label{fig:fig_method_flowchart}
	\end{center}
 \end{figure}

The key components of the optimization loop are illustrated in \Cref{fig:fig_method_flowchart}. 

\begin{algorithm}[t]
\caption{Finite Element Analysis}
\label{alg:fea}
\begin{algorithmic}[1]
    \Require Mesh, Dirichlet and Neumann BCs, effective $G$ and $\chi$ distribution
    \Comment{\cref{eq:eff_G,eq:eff_chi}}
         
    \Require $\mu_{\mathrm{dry}}$, $\mu_{\mathrm{wet}}$, $\mu^0$; load schedule $\{\alpha^{[k]}\}_{k=1}^{N_s}$ where $\alpha^{[k]} = (k/N_s)^{\beta}$
    \Comment{$\beta \in (0,1)$, $\beta \approx 0.05$ in our expts.}
         
    \Require NR tolerance $\epsilon_{nr}$, max NR iterations $I_{nr}$

    \State Initialize $\bm{u}^{[0]} \leftarrow \bm{0}$

    \For{$k = 1$ \textbf{to} $N_s$} \Comment{Load stepping}

        \State $\mu^{[k]} \leftarrow (1 - \alpha^{[k]})\,\mu_{\mathrm{dry}} + \alpha^{[k]}\,\mu_{\mathrm{wet}}$
       \Comment{Load step chemical potential}
           
        \State Scale $\bar{\bm{t}}$ and non-homogeneous $\bar{\bm{u}}$ by $\alpha^{[k]}$
        \Comment{Load step BCs}
           
        \State $\bm{u}^{(0)} \leftarrow \bm{u}^{[k-1]}$; \;  $\bm{u}^{(0)}\big|_{\Gamma_D} \leftarrow \alpha^{[k]}\,\bar{\bm{u}}$
        \Comment{Warm start from previously load step (and impose BC)}
        
        \State $l \leftarrow 0$,\quad $R_{\mathrm{norm}} \leftarrow \infty$
        \Comment{Init. NR counter and residual}
        
         \While{$R_{\mathrm{norm}} > \epsilon_{nr}$ \textbf{and} $l < I_{nr}$}
         
        \State $\bm{f}_{\mathrm{int}} \leftarrow \bm{0}$,\quad $\bm{K}_T \leftarrow \bm{0}$
        \Comment{Init. internal force and tangent stiffness}
        
        \For{each integration point}
        
            \State Compute deformation gradient $\bm{F}$ from 
                   $\bm{u}^{(l)}$
            
            \State Compute in-plane right Cauchy-Green tensor $\bm{C}_{2D} = \bm{F}^\top\bm{F}$
            \Comment{plane stress assumption}
                   
            \State Compute in-plane area stretch $J_{2D} = \sqrt{\det \bm{C}_{2D}} = \lambda_1\lambda_2$
            \Comment{In-plane principal stretches product}
                   
            \State $\phi \leftarrow \textsc{Bisection}(J_{2D},\, G,\, \chi,\, \mu^{[k]},\, \mu^0,\, \tilde{\Omega},\, T)$
            \Comment{solves \cref{eq:flory_rehner} via bisection}
            
            \State Compute $J \leftarrow 1/\phi$,\quad $\lambda_3^2 \leftarrow (J / J_{2D})^2$
           \Comment{Molecular incompressibility; \cref{sec:method_governingEquations}}
            
           \State Compute $I_1 \leftarrow \mathrm{tr}(\bm{C}_{2D}) + \lambda_3^2$
           \Comment{Eq.~\eqref{eq:I1}}
            
            \State Compute $\psi$ via Eq.~\eqref{eq:free_energy}; obtain PK-1 stress $\bm{P}$ and tgt. stiffness $\bm{K}_T^e$ via AD of $\psi$

            \State Accumulate $\bm{f}_{\mathrm{int}}$, $\bm{K}_T$
        
        \EndFor
        
        \State $\bm{R}^{(l)} \leftarrow \bm{f}_{\mathrm{ext}} - \bm{f}_{\mathrm{int}}$
       \Comment{Eq.~\eqref{eq:residual}}
        
        \State Apply Dirichlet BCs to $\bm{R}^{(l)}$, $\bm{K}_T$
        
        \State Solve $\bm{K}_T\,\delta\bm{u} = -\bm{R}^{(l)}$
        \Comment{PETSc sparse direct solver}
        
        \State $\bm{u}^{(l+1)} \leftarrow  \bm{u}^{(l)} + \delta\bm{u}$
       
        \State $R_{\mathrm{norm}} \leftarrow \|\bm{R}^{(l+1)}\|$;\quad
               $l \leftarrow l+1$
    \EndWhile
    \State $\bm{u}^{[k]} \leftarrow \bm{u}^{(l)}$
           \Comment{Store converged solution at load step $k$}
\EndFor
\State \Return $\bm{u}^\star \leftarrow \bm{u}^{[N_s]}$
\end{algorithmic}
\end{algorithm}

\section{Numerical Studies}
\label{sec:expts}

In this section, we conduct several numerical studies to illustrate the proposed framework. All experiments are conducted on a MacBook M3 Pro using the JAX library \cite{bradbury2018jax} in Python. Unless otherwise specified, the default parameters for all numerical examples are as follows:

\begin{itemize}

   \item \textbf{Neural network} : A network with $2$ hidden layers; with $40$ neurons in each layer; totaling $17,969$ trainable parameters is used.

   \item \textbf{Constraints} : We enforce a grayness constraint (\Cref{eq:grayness_constraint}) governed by a continuation strategy, where the slack variable $\xi$ is initialized to $2$ and linearly decremented by $5 \times 10^{-2}$ per iteration to a final value of $5 \times 10^{-2}$. The SIMP penalty parameter starts from 1 and is increased by 0.05 till it reaches a value of 3.
   
   \item \textbf{Optimizer} : The Adam optimizer \cite{kingma2014adam} with a learning rate of $5 \times 10^{-3}$ is used. To improve stability, a gradient clip with a norm threshold of $1$ is used.

    \item \textbf{Convergence} : The optimization is terminated either till a maximum of $250$ iteration or when the loss residual $ \Delta \mathcal{L} \leq 10^{-3}$.

\end{itemize}

\subsection{Case \Romannum{1}: Design for Shape Morphing}
\label{sec:expts_shapeMorphing}

We begin by considering the design of structures that morph to a target shape upon swelling. The underpinning principle is that bonding a swellable gel layer to a non-swelling elastomer layer produces differential volumetric expansion at the interface upon solvent exposure, driving the composite to bend out of plane. This is the soft-matter analog of the classical bimetallic strip \cite{Timoshenkobimetal}, and has been exploited to produce programmed shape changes in stimuli-responsive structures for applications in soft robotics \cite{jiao2022programmableSoftRoboticsHydrogel}, biomedical devices \cite{zhuo2020ComplexMultiphase}, and 4D-printed designs \cite{sydneygladman2016Biomimetic4d}.

\begin{figure}[]
 	\begin{center}
		\includegraphics[scale=0.35,trim={0 0 0 0},clip]{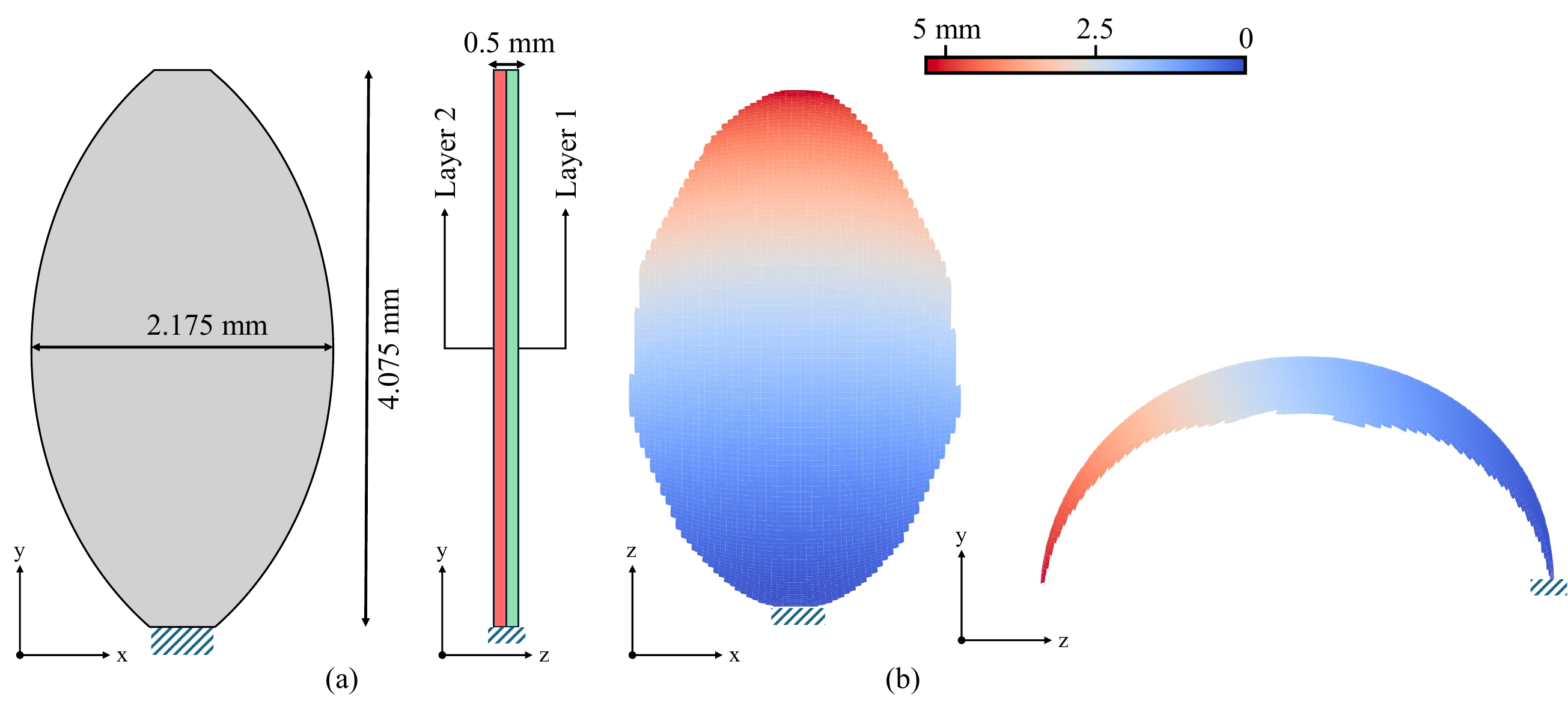}
 		\caption{Swelling-induced bending of a uniform bilayer petal-shaped structure. (a) Geometry of the bilayer in the undeformed configuration (left: $x$-$y$ plane view; right: $y$-$z$ plane cross-section showing the gel and elastomer layers). (b) Deformed configurations under solvent exposure, colored by displacement magnitude.}
    \label{fig:fig_result_petal_bilayer_swelling}
	\end{center}
 \end{figure}

For illustration, consider a simple petal-shaped bilayer, as illustrated in \Cref{fig:fig_result_petal_bilayer_swelling}(a), with uniform gel and elastomer layers bonded along the thickness. Upon immersion in solvent, the gel swells while the elastomer resists, producing the large out-of-plane bending shown in \Cref{fig:fig_result_petal_bilayer_swelling}(b). The tip displacement reaches approximately $5$~mm, exceeding the semi-major axis of the structure, demonstrating the large deformation regime captured by the forward model.

To achieve non-trivial, spatially programmed deformation profiles, one requires beyond uniform layer compositions. We pose this as the shape-morphing optimization problem defined in \Cref{eq:objective_morphing}, where the objective is to minimize the mean-squared error between the realized displacement field and a prescribed target. This is formulated as a two-layer design problem, where the material compositions of the top and bottom layers are treated as independent design fields. We consider a target deformation profile as motivated in \Cref{fig:fig_method_graphicalAbstract}. The structure is modeled using shell elements, where the material at Gauss points along the through-thickness direction attributed independently, yielding separate top and bottom-layer design fields. The gel and elastomer parameters are set to $\chi_{\mathrm{gel}} = 0.2$, $G_{\mathrm{gel}} = 1 \times 10^{6}$~Pa and $\chi_{\mathrm{elast}} = 5$, $G_{\mathrm{elast}} = 5 \times 10^{7}$~Pa, respectively. A solvent with molar volume $\tilde{\Omega} = 1.8 \times 10^{-5}$~m$^{3}$~mol$^{-1}$ at a temperature $T = 298$~K is assumed. Furthermore, chemical bath potentials with $\mu_{\mathrm{dry}} = -1 \times 10^{5}$~J~mol$^{-1}$, $\mu^{0} = 0$, and $\mu_{\mathrm{wet}} = -100$~J~mol$^{-1}$ are assumed.

\begin{figure}[]
 	\begin{center}
		\includegraphics[scale=0.45,trim={0 0 0 0},clip]{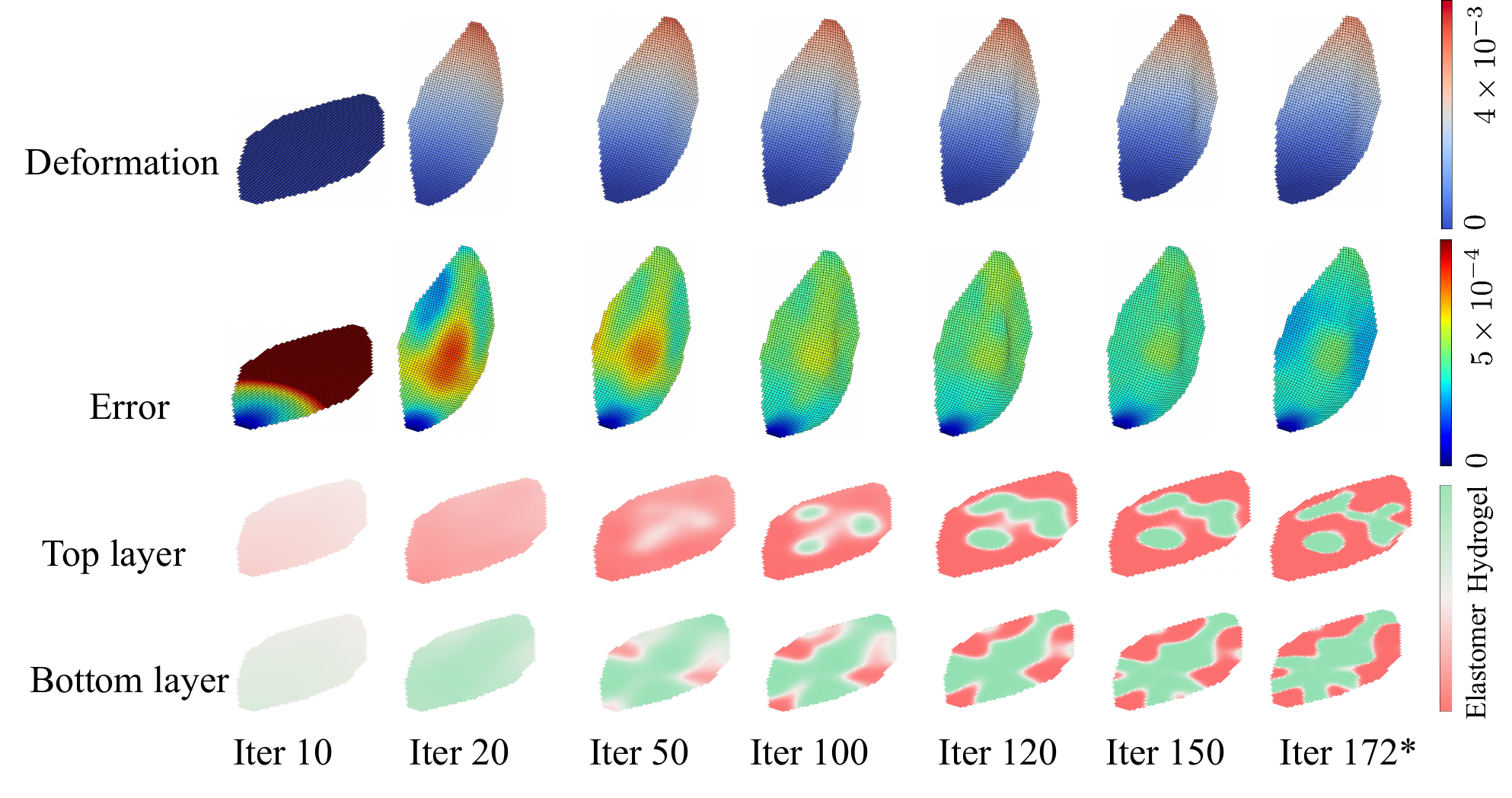}
 		\caption{Evolution of the deformed configuration, the error from the target distribution, and the material distribution of the top and bottom layers during the optimization.}
    \label{fig:fig_result_convg_shapeMorphingLeaf}
	\end{center}
 \end{figure}

The evolution of the deformation profile, the error with respect to the target deformation, and the material distributions of the two layers across iterations are shown in \Cref{fig:fig_result_convg_shapeMorphingLeaf}. The optimization significantly reduces the objective function, with the mean squared error (MSE) decreasing to $1.42 \times 10^{-4}$ over 172 iterations. As shown in the 'Error' row of \Cref{fig:fig_result_convg_shapeMorphingLeaf}, the spatial discrepancy between the realized and target configurations is progressively minimized, with the final peak nodal error reduced to $2.14 \times 10^{-4}$ m.

\subsection{Case \Romannum{2}: Design of a Swelling Actuated Inverter}
\label{sec:expts_inverter}

 \begin{figure}[]
 	\begin{center}
		\includegraphics[scale=0.25,trim={0 0 0 0},clip]{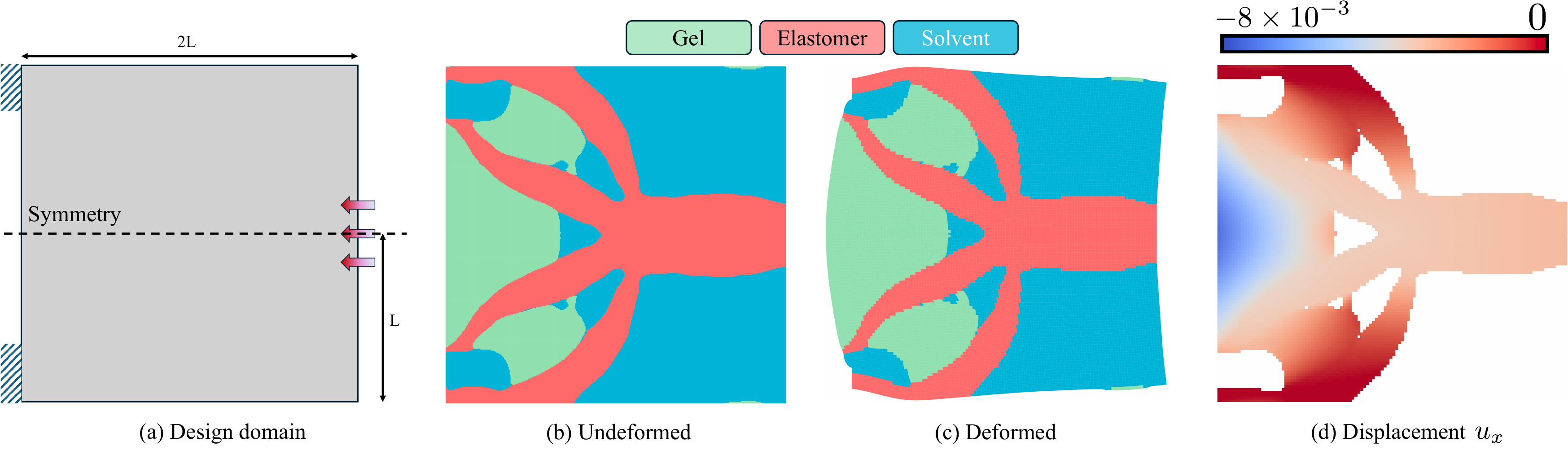}
 		\caption{Swelling-actuated mechanical inverter. (a) Design 
  domain and boundary conditions. (b) Undeformed optimized topology. (c) Deformation upon solvent exposure. The gel swells and compresses the elastomer arms, deforming of the output port in the $-X$ direction.}
    \label{fig:fig_result_inverter}
	\end{center}
 \end{figure}

As a second design example, we consider a swelling-actuated mechanical inverter \cite{kumar2020TOCompliantFluid,zhu2020mechanismDesign}, in which the elastomeric skeleton redirects gel expansion into an output displacement opposing the natural outward expansion. Such compliant mechanisms are relevant to designing soft grippers \cite{shepherd2011MultigaitSoft} and biomedical devices \cite{wehner2016IntegratedDesign}. The design domain, and boundary conditions are illustrated in \Cref{fig:fig_result_inverter}(a). The upper half is optimized assuming a line symmetry, with the output port at the right edge. The reaction force in the $-X$ direction at this port (push force) is the quantity of interest for our optimization (\Cref{eq:objective_blocked_force}). The material and solvent parameters are identical to those in \Cref{sec:expts_shapeMorphing}. The output nodes are fixed during each forward solve with the push force is maximized.

\begin{figure}[H]
 	\begin{center}
		\includegraphics[scale=0.4,trim={0 0 0 0},clip]{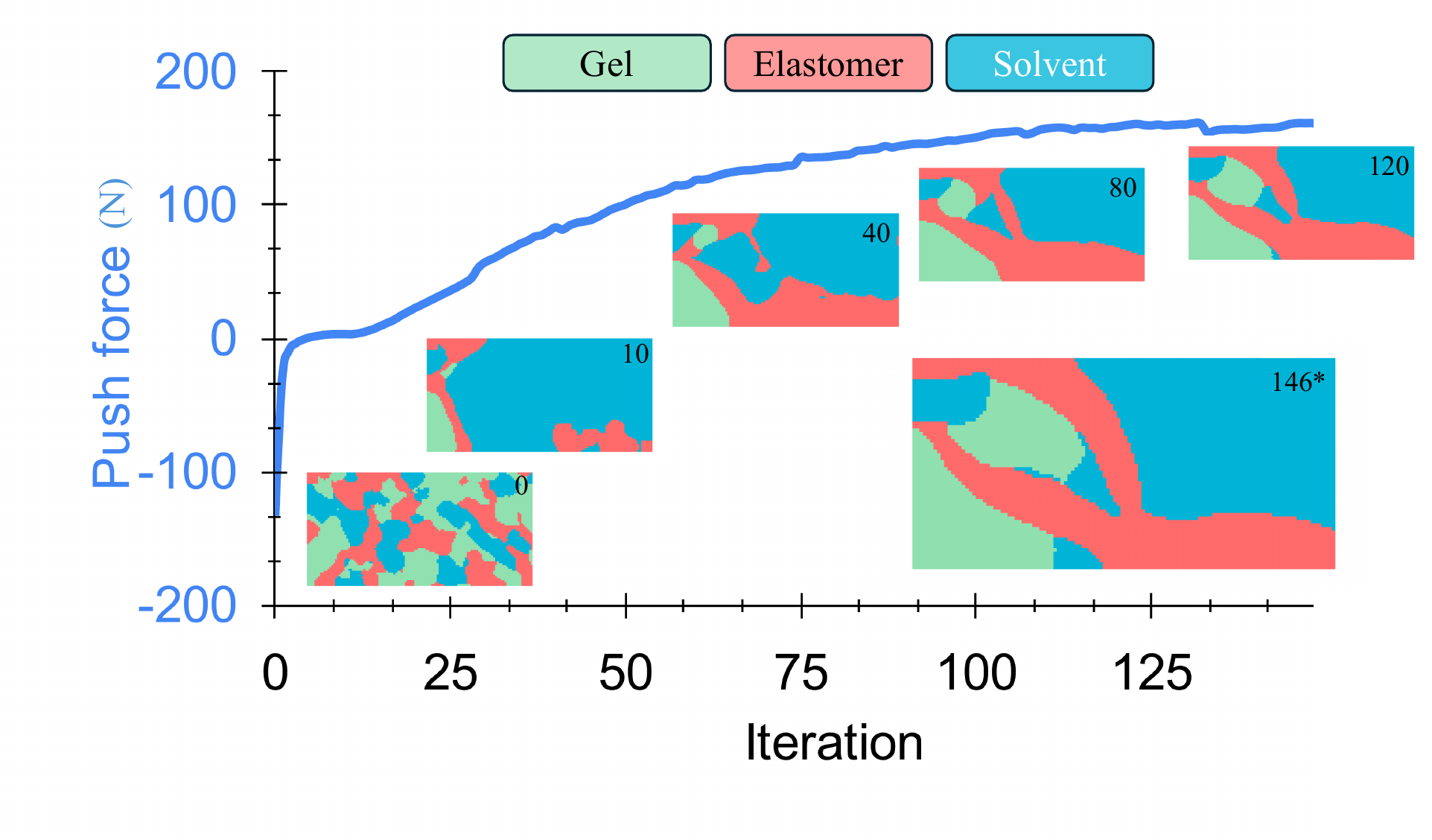}
 		\caption{ Convergence history of the push force at the output port over the course of the optimization. Inset snapshots show the material distribution at selected iterations.}
    \label{fig:fig_result_inverter_convg_panels}
	\end{center}
 \end{figure}

The convergence history is shown in \Cref{fig:fig_result_inverter_convg_panels}. The push force begins at  $\approx -150$~N, as the initial disordered topology drives the output port in the natural outward direction. The push force increases and converges to $+153$~N after $146$ iterations, with the intermediate snapshots showcased tracing the evolution of the design.

The optimized topology in the undeformed and deformed configurations is shown in \Cref{fig:fig_result_inverter}(b,c). Two elastomer arms extend from the fixed supports and connect to a central spine at the output port, with gel seated in the enclosed regions. Upon swelling, the gel exerts compressive forces on the arms, which being anchored at the left edge transmit this load inward along the spine, producing the desired $-X$ displacement at the output. Observe that the elastomer undergoes no volumetric swelling; its deformation is purely mechanical, driven by load transfer across the gel-elastomer interfaces.

\subsection{Case \Romannum{3}: Design of Organogel-Hydrogel Composites}
\label{sec:expts_organohydrogels}

In this experiment, we extend the framework to structures composed of three phases: hydrogel, organogel, and elastomer. The hydrogel swells in water but remains inert to organic solvents, while the organogel swells in organic solvents but remains inert to water. This solvent-selective response has been showcased in the literature to produce bidirectional shape changes in stimuli-responsive actuators and chemical-sensing structures\cite{zhuo2020organohydrogels,lee2025organogelhydrogel}, motivating applications in environmental monitoring and solvent-selective drug release \cite{zhuo2020organohydrogels,boroomand2025opticalSensingHydrogel}.

\begin{figure}[]
 	\begin{center}
		\includegraphics[scale=0.35,trim={0 0 0 0},clip]{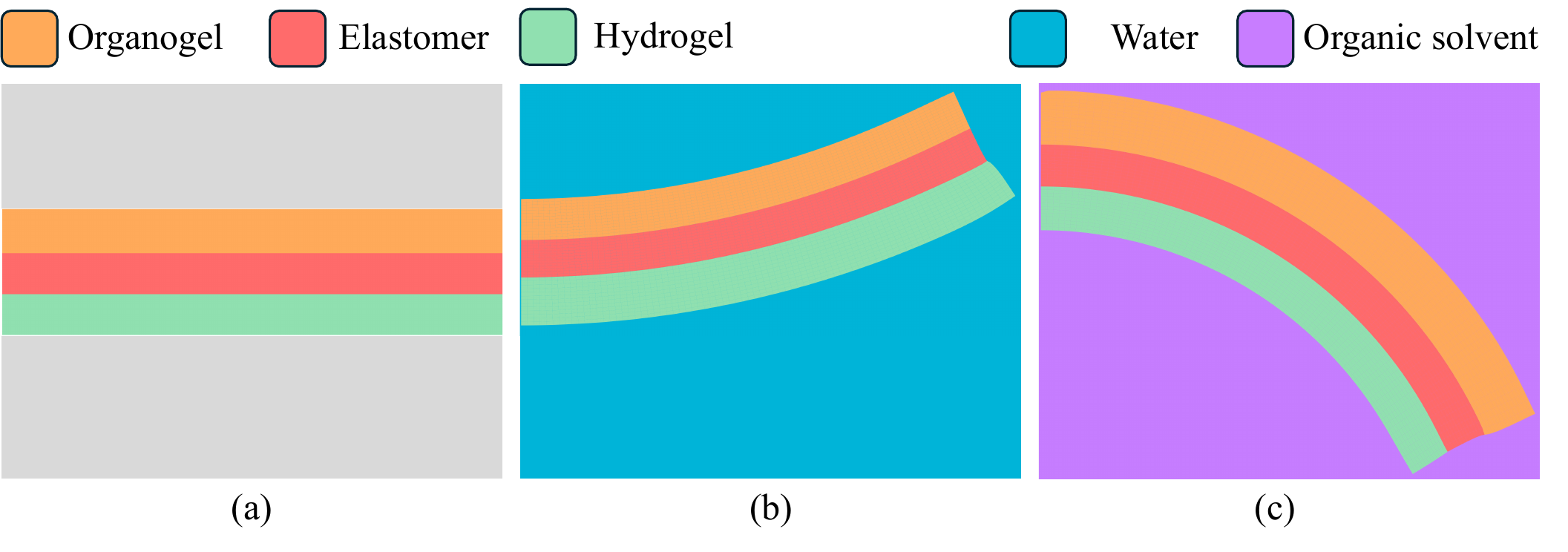}
 		\caption{Solvent-selective actuation in a uniform trilayer strip. (a) Undeformed configuration, composed of an organogel (top layer), elastomer (middle layer), and hydrogel (bottom layer). (b) Deformed configuration upon immersion in water: the hydrogel swells and drives the strip upward. (c) Deformed configuration upon immersion in organic solvent: the organogel swells and drives the strip downward.}
    \label{fig:fig_result_organohydrogel_simulation}
	\end{center}
 \end{figure}

We present a simplified illustration of this solvent-specific response in \Cref{fig:fig_result_organohydrogel_simulation}. A uniform trilayer composite composed of organogel, elastomer, and hydrogel layers is shown in its undeformed state in \Cref{fig:fig_result_organohydrogel_simulation}(a). Upon immersion in water (\Cref{fig:fig_result_organohydrogel_simulation}(b)), the hydrogel swells while the organogel remains inert, driving the strip upward. Upon immersion in organic solvent (\Cref{fig:fig_result_organohydrogel_simulation}(c)), the organogel swells while the hydrogel remains inert, producing the opposing downward deflection. This capability, whereby the structure produces distinct actuated states under distinct chemical environments, motivates its use in a multi-load TO setting.

To this end, we consider the same design domain and boundary conditions as in \Cref{sec:expts_inverter} (\Cref{fig:fig_result_inverter}(a)), now with three material phases (and void) available to the optimizer. The multi-load problem requires the structure to function as an inverter at the output port, displacing it in the $-X$ direction upon exposure to water, while functioning as a pusher, displacing the output port in the $+X$ direction, upon exposure to organic solvent. The Flory-Huggins parameters are $\chi = 0.15$ for affine pairs (hydrogel in water; organogel in organic solvent) and $\chi = 5.0$ for all remaining material-solvent combinations, including the elastomer and void phases. The shear modulus are $G_{\mathrm{elastomer}} = 1 \times 10^{7}$~Pa, $G_{\mathrm{hydrogel}} = 1 \times 10^{6}$~Pa, and $G_{\mathrm{organogel}} = 2 \times 10^{6}$~Pa. The two solvents are water with molar volume $\tilde{\Omega}_{\mathrm{water}} = 1.8 \times 10^{-5}$~m$^3$~mol$^{-1}$ and an organic solvent with $\tilde{\Omega}_{\mathrm{organic}} = 1 \times 10^{-4}$~m$^3$~mol$^{-1}$, both at $T = 298$~K and with the same chemical potential schedule as in \Cref{sec:expts_inverter}. The output nodes are held fixed in each load case; the design objective is to maximize the reaction force at the output port in the $+X$ direction (\Cref{eq:objective_blocked_force}) under organic solvent exposure, subject to the constraint that the reaction force in the $-X$ direction under water exposure exceeds $20$~N. A volume constraint of $0.4$ is imposed on the total solid phase fraction.

\begin{figure}[]
 	\begin{center}
		\includegraphics[scale=0.45,trim={0 0 0 0},clip]{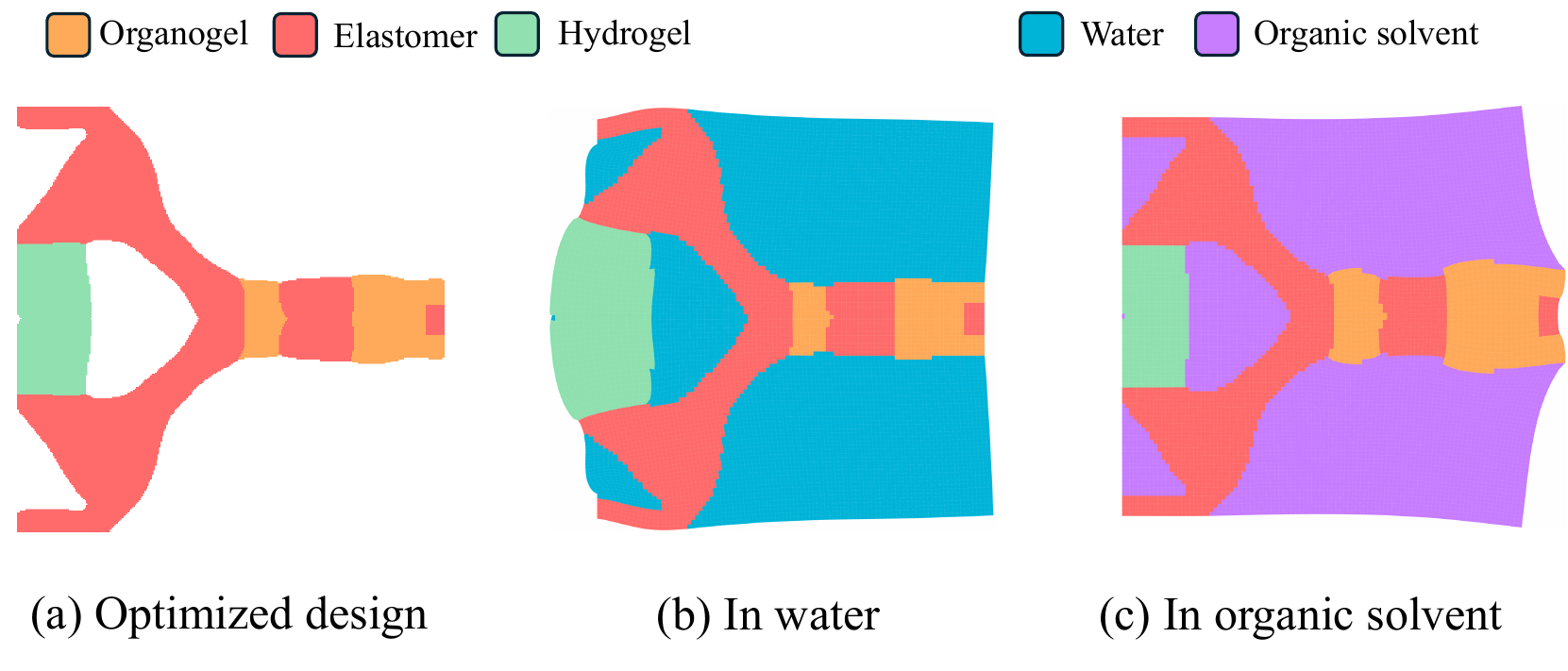}
 		\caption{Optimized multi-stimuli organo-hydrogel actuator. (a) Optimized design. (b) Deformed configurations upon immersion in water, where the output port displaces inward ($-X$), and (c) in organic solvent, where the 
  output port displaces outward ($+X$).}
    \label{fig:fig_result_organohydrogelInverter}
	\end{center}
 \end{figure}

The optimized design and its deformed configurations are shown in \Cref{fig:fig_result_organohydrogelInverter}. In the optimized topology (\Cref{fig:fig_result_organohydrogelInverter}(a)), the hydrogel is placed in two large pockets flanking the left fixed supports, while the organogel occupies a central horizontal band connecting directly to the output spine; the elastomer forms the enclosing structural frame. This spatial segregation reflects the functional role of each phase: in water, the hydrogel pockets swell and compress the elastomer arms inward, transmitting a net $-X$ force to the output port (\Cref{fig:fig_result_organohydrogelInverter}(b)); in organic solvent, the organogel band swells and extends the central spine outward, producing the desired $+X$ displacement at the output port (\Cref{fig:fig_result_organohydrogelInverter}(c)).

\subsection{Case \Romannum{4}: Design with Anisotropic Hydrogel}
\label{sec:expts_anisotropichydrogels}

 \begin{figure}[]
 	\begin{center}
		\includegraphics[scale=0.45,trim={0 0 0 0},clip]{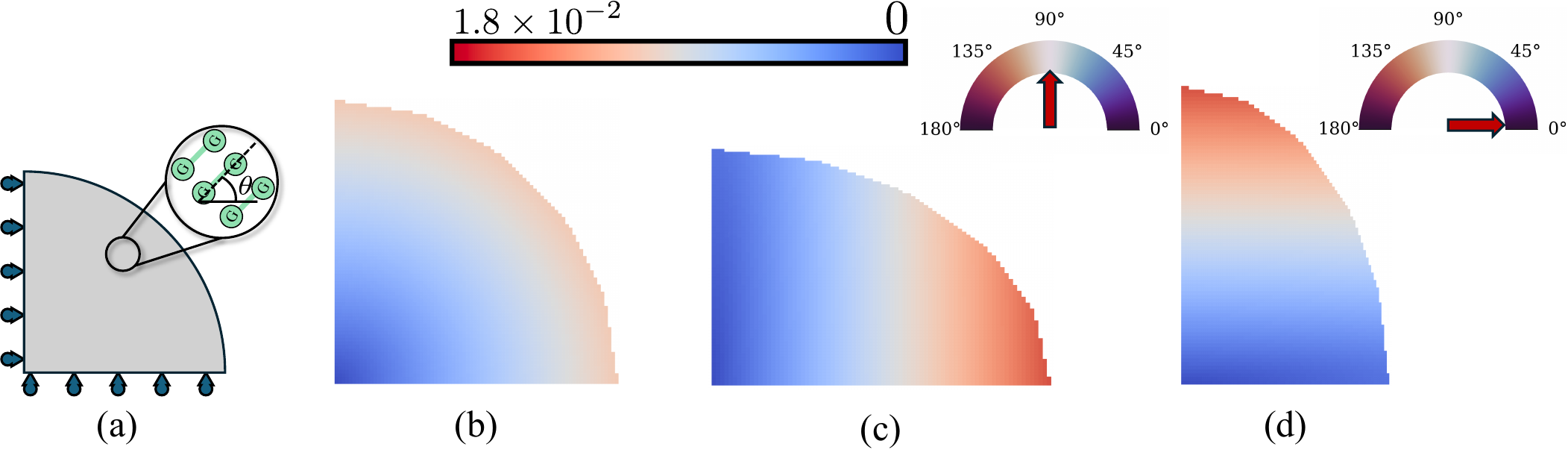}
 		\caption{Effect of fiber orientation on anisotropic swelling of hydrogel. (a) Geometry and boundary conditions (b) isotropic gel (no fiber), (c) fiber aligned at $90^{\circ}$, and (d) fiber aligned at $0^{\circ}$.}
    \label{fig:fig_result_aniso_swelling}
	\end{center}
 \end{figure}
 
In this section, we consider the design of programmable structures composed of anisotropic hydrogels. The experiments presented hitherto have exploited spatial heterogeneity in the gel and elastomer distribution to program bending-dominated deformation profiles. However, deformation modes involving twist remain inaccessible to isotropic multi-material designs, since the swelling eigenstrain $\bm{F}^s = \phi^{-1}\mathbf{1}$ is isotropic by construction, restricting the through-thickness stress integral to bending moments while leaving the twisting moment zero. Introducing anisotropy overcomes this restriction. This mechanism has direct biological precedent in plant organ actuations \cite{sleboda2023anisotropyPlant}, and has been exploited in biomimetic 4D printing of hydrogel composites \cite{sydneygladman2016Biomimetic4d}. To model anisotropic hydrogels, the strain energy density in \Cref{eq:free_energy} is augmented with a tension-only fiber term following a Holzapfel-Gasser-Ogden (HGO) model \cite{gasser2006HGO}:

\begin{equation}
\Psi = \underbrace{\frac{G}{2} (I_1 - 3 - 2 \ln J)}_{\Psi_{\text{gel}}} + \underbrace{\frac{\eta_f}{2} \langle I_4 - 1 \rangle_{+}^{2}}_{\Psi_{\text{aniso}}}
\label{eq:aniso_gel_psi}
\end{equation}

where $a_0 = [\cos \theta, \sin \theta]$ is the local unit fiber direction vector, $I_4 =  a_0 . C . a_0$, $\langle x \rangle_{+} = \max(0, x)$, and $\eta_f$ is the fiber stiffness. To illustrate the effect of fiber orientation on swelling, we consider a (quarter symmetric) circular plate of uniform hydrogel as shown in \Cref{fig:fig_result_aniso_swelling}(a). Without fibers, \Cref{fig:fig_result_aniso_swelling}(b) shows isotropic swelling. With fibers at $90^{\circ}$ and $0^{\circ}$, \Cref{fig:fig_result_aniso_swelling}(c) and (d) show that the peak displacement is redirected toward the direction perpendicular to the fiber, since molecular incompressibility forces the suppressed expansion along $\bm{a}_0$ to be compensated by larger stretch transverse to it.

 \begin{figure}[]
 	\begin{center}
		\includegraphics[scale=0.4,trim={0 0 0 0},clip]{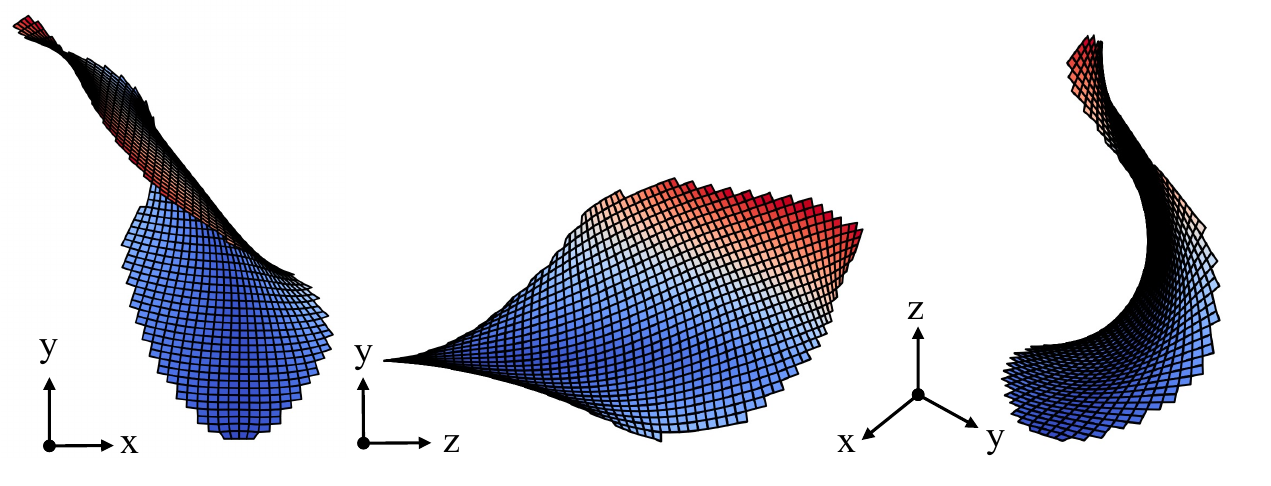}
 		\caption{Target deformation profile for the anisotropic shape-morphing}
    \label{fig:fig_results_anisoTargetSwelling}
	\end{center}
 \end{figure}

\begin{figure}[]
 	\begin{center}
		\includegraphics[scale=0.4,trim={0 0 0 0},clip]{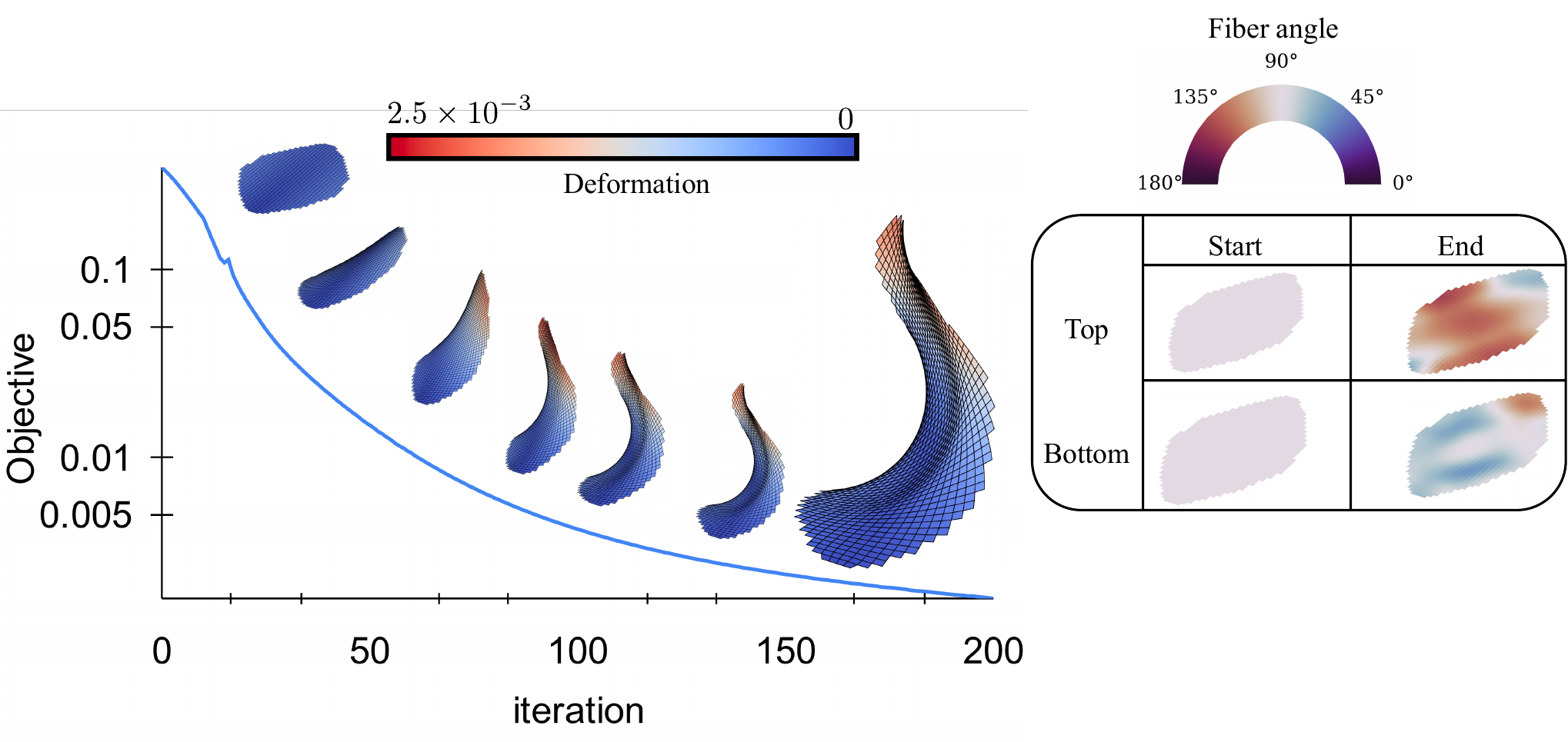}
 		\caption{Optimization convergence for the anisotropic shape-morphing experiment. Left: objective (MSE) versus iteration on a logarithmic scale with instances of deformed configurations as inset. Right: initial and final fiber orientation fields for the top and bottom layers.}
    \label{fig:fig_results_fibAnisoOptResult}
	\end{center}
 \end{figure}

We now consider two design scenarios with anisotropic hydrogels. All material and solvent parameters are identical to those in \Cref{sec:expts_inverter} across both experiments, with the additional fiber stiffness parameter $\eta_f = 5G_{\text{gel}}$. In the first experiment, we revisit the petal-shaped domain of \Cref{sec:expts_shapeMorphing} and treat the fiber orientation field $\theta(\bm{x})$ at the top and bottom layers as the sole design variables, with the material distribution fixed. The synthetic target deformation profile shown in \Cref{fig:fig_results_anisoTargetSwelling} requires significant spatial variation in orientation for recovery. The fiber angle at each layer is parameterized by a neural network that maps spatial coordinates of the through thickness integration points at the top and bottom to $\theta$, with the output activated by a sigmoid and scaled to $[0, \pi)$ to span all distinct fiber orientations. The optimization objective is the mean-squared displacement error between the simulated and target fields, evaluated at the shell nodes as in \Cref{eq:objective_morphing}. The convergence history, alongside the deformed configurations and fiber orientation fields at selected iterations, is shown in \Cref{fig:fig_results_fibAnisoOptResult}. Here, the optimization converges over 211 iterations, reducing the objective function to $1.2 \times 10^{-3}$. The framework accurately captures the localized anisotropic responses, with the final realized state matching the prescribed target configuration.

\begin{figure}[]
 	\begin{center}
		\includegraphics[scale=0.3,trim={0 0 0 0},clip]{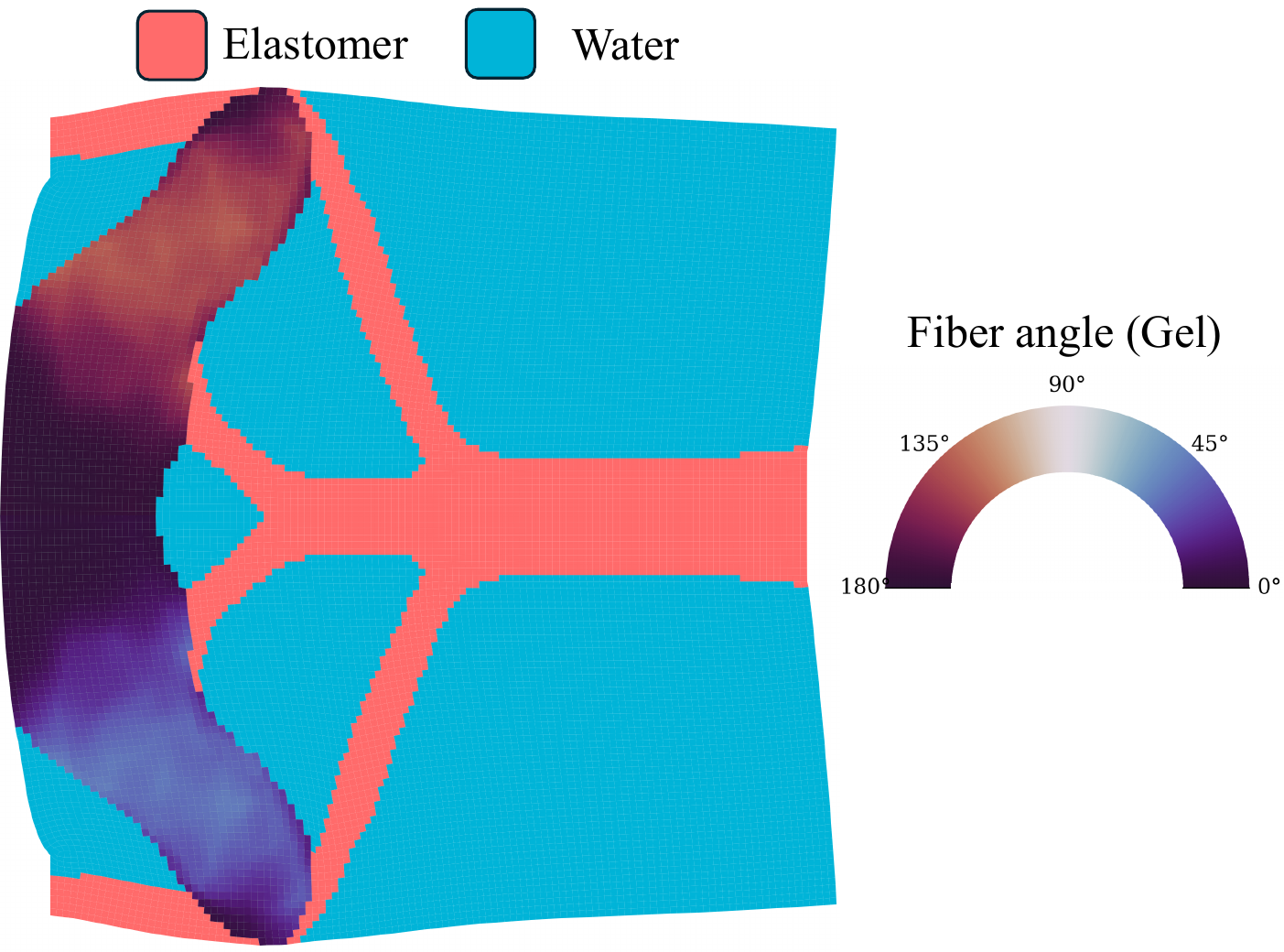}
 		\caption{Optimized anisotropic inverter.}
    \label{fig:fig_results_inverterAniso}
	\end{center}
 \end{figure}

In the second experiment, we extend the swelling actuated inverter of \Cref{sec:expts_inverter} to the anisotropic setting, now optimizing concurrently over both the material distribution $\bm{\rho}(\bm{x})$ and the fiber orientation field $\theta(\bm{x})$. Anisotropy is attributed solely to the hydrogel phase, with the fiber stiffness interpolated via the SIMP scheme as $\eta_f(\bm{x}) = \rho_g^p(\bm{x})\,\eta_f^{\text{gel}}$, so the elastomer phase remains isotropic throughout. The optimized material distribution and fiber orientation field are shown in \Cref{fig:fig_results_inverterAniso}. The elastomeric skeleton recovers a topology similar to that of \Cref{sec:expts_inverter}, with two arms anchored at the fixed supports transmitting load to the output spine. Within the gel regions, the optimized fiber orientations trend predominantly transverse to the local swelling expansion direction, consistent with the directional stiffening demonstrated in the quarter-plate example. This experiment demonstrates that the framework naturally accommodates the concurrent optimization of topology and fiber orientation under the same differentiable pipeline.

\section{Conclusion}
\label{sec:conclusion}

This work proposes GELATO, a framework for the design of programmable \underline{g}el-\underline{ela}stomer composites using \underline{t}opology \underline{o}ptimization . The composite couples an active, swellable hydrogel with a passive elastomer, allowing complex large-deformation profiles to be programmed through optimized material distribution. The framework utilizes a nonlinear neo-Hookean constitutive model based on Flory-Rehner theory. Furthermore, a coordinate-based neural network parameterizes the design by mapping spatial coordinates to material pseudodensities, decoupling the topology from the finite element discretization. Finally, the framework is implemented in an end-to-end differentiable manner, with sensitivities computed via reverse-mode automatic differentiation and implicit differentiation of the nonlinear equilibrium equations. Several numerical experiments demonstrate the utility of the approach, including programmable shape-morphing structures and compliant inverters. The framework is further extended to organogel-hydrogel composites, enabling multi-stimuli responsiveness across chemically diverse media, and to anisotropic hydrogels, where the local fiber orientation is optimized concurrently with the topology, adding a further degree of programmability.

Together, these capabilities enable automated inverse design of swellable structures targeting arbitrary shape changes, a task previously tractable only through manual forward-simulation sweeps. Several directions remain open to extend this further. The fidelity of the material models can be improved by incorporating finite-strain potentials such as Gent or Arruda-Boyce \cite{gent1996Rubber,arruda1993Rubber} at extreme stretches. Furthermore, implementing thermo-chemo-mechanically coupled formulations \cite{chester2011thermalSwelling,brunner2024chemoThermalHydrogel} and accounting for interfacial debonding \cite{tian2018adhesionHydrogelElastomer} would narrow the sim-to-reality gap. Modeling transient and viscoelastic effects is key to capturing rate-dependent kinetics and enabling the design of 4D-printed structures that follow prescribed actuation trajectories \cite{bouklas2015nonlinearHydrogelTransientSolventDiffusion}. Extending the framework to encompass multi-physics stimuli, such as pH-responsive \cite{marcombe2010pHHydrogel} and magnetically responsive hydrogels, would broaden the accessible design space. Optimizing for variable hydrogel crosslink density \cite{hoti2021CrossLinkDensityHydrogel} could facilitate in realizing functionally graded materials, introducing an additional degree of programmability. Incorporating robust optimization formulations \cite{schevenels2011robustTO} and printability requirements is essential for the physical realization of these designs \cite{vatanabe2016ManufCons}. Finally, scaling the framework to 3D volumetric problems using GPU-accelerated nonlinear solvers \cite{alexandersen2026largeTransientTO,macklin2022NvidiaWarp} would enable the automated design of complex programmable hydrogels. Ultimately, GELATO provides a scalable, differentiable platform for the inverse design of swellable programmable material systems with direct applicability to soft actuators, biomedical devices, and microfluidic systems.

\section*{Appendix: Implicit Differentiation}
\label{sec:appendix_a}

We employ gradient-based optimization, driven by the Adam optimizer, to minimize the loss $\mathcal{L}$ with respect to the network weights $\bm{w}$ (\Cref{eq:loss}). While the chain rule can be trivially applied automatically for most of the computational graph, two terms involve the sensitivities via implicit fields: the displacement $\bm{u}$, which satisfies the discrete equilibrium $\bm{R}(\bm{u};\bm{w}) = \bm{0}$
(\Cref{eq:residual}) resolved via Newton-Raphson iteration, and the polymer volume fraction $\phi$, which satisfies the Flory-Rehner equation $\mathcal{F}(\phi;\bm{u},G,\chi) = 0$ (\Cref{eq:flory_rehner}) resolved via bisection. Note that both involve iterative root-finding procedures: neither
field admits a closed-form expression in $\bm{w}$. A naive application of AD would unroll the derivative computation graph across every iterate of both solvers, incurring memory and compute costs that scale with the iteration count. We instead apply the Implicit Function Theorem (IFT) uniformly to both.

\paragraph{Implicit Function Theorem.}
For a converged solution $s^\star$ satisfying $\mathcal{H}(s^\star;\bm{\theta}) = 0$, the IFT \cite{blondel2022implicitdiff} states that, provided $\partial\mathcal{H}/\partial s$ is invertible at the converged state:

\begin{equation}
  \frac{ds^\star}{d\bm{\theta}}
  = -\left(\frac{\partial\mathcal{H}}{\partial s}\right)^{-1}
     \frac{\partial\mathcal{H}}{\partial\bm{\theta}}.
\end{equation}

This replaces differentiation through an arbitrary iteration history with a single evaluation of two partial derivatives at $s^\star$, and applies to any root-finding scheme regardless of the specific iteration strategy. We instantiate this for each solver in turn.

For the Flory-Rehner equation, $\phi^\star$ satisfies the scalar residual $\mathcal{F}(\phi^\star;\bm{u},G,\chi) = 0$ at each quadrature point. Observe that $\mathcal{F}$ depends on $\bm{w}$ only indirectly, through the material properties $G(\bm{w})$ and $\chi(\bm{w})$ supplied by the network, and through
the deformation state $\bm{u}$. Setting $s \equiv \phi^\star$
and $\bm{\theta} \equiv (\bm{u}, G, \chi)$:

\begin{equation}
  \frac{d\phi^\star}{d\bm{\theta}}
  = -\left(\frac{\partial\mathcal{F}}{\partial\phi}\right)^{-1}
     \frac{\partial\mathcal{F}}{\partial\bm{\theta}}.
\end{equation}

Note that $\partial\mathcal{F}/\partial\phi$ is a non-zero scalar for all physically realizable states, so the inversion reduces to a scalar division. Since $\mathcal{F}$ is a smooth, closed-form expression (Equation~\eqref{eq:flory_rehner}), both partial derivatives are obtained directly by AD of Equation~\eqref{eq:flory_rehner}, without any iteration. The remaining chain from $\bm{\theta}$ to $\bm{w}$ is then handled automatically by AD through the network.

Similarly, the converged displacement $\bm{u}^\star$ satisfies $\bm{R}(\bm{u}^\star;\bm{w}) = \bm{0}$, where the residual internally calls the bisection solver at each quadrature point. Setting $s \equiv \bm{u}^\star$ and $\bm{\theta} \equiv \bm{w}$:

\begin{equation}
  \frac{d\bm{u}^\star}{d\bm{w}}
  = -\bm{K}^{-1}
    \left.\frac{\partial\bm{R}}{\partial\bm{w}}\right|_{\bm{u}^\star},
\end{equation}

where $\bm{K} = \left.\partial\bm{R}/\partial\bm{u} \right|_{\bm{u}^\star}$ is the tangent stiffness matrix at the converged state. Observe that when AD differentiates $\bm{R}$ with respect to $\bm{w}$ to form the right-hand side, it traces through the residual evaluation and encounters the bisection call; since the bisection solver already carries its IFT rule, the gradient propagates implicitly through the coupled swelling response without further intervention. In practice, forming the full sensitivity matrix $d\bm{u}^\star/d\bm{w} \in \mathbb{R}^{n_u \times n_w}$ is infeasible for large systems. We instead compute its action on a tangent vector $\bm{v}$ via a Jacobian-vector product (JVP):

\begin{equation}
  \frac{d\bm{u}^\star}{d\bm{w}} \cdot \bm{v}
  = -\bm{K}^{-1}
    \left(
      \left.\frac{\partial\bm{R}}{\partial\bm{w}}\right|_{\bm{u}^\star}
      \cdot \bm{v}
    \right),
\end{equation}

where $\left.\partial\bm{R}/\partial\bm{w}\right|_{\bm{u}^\star} \cdot \bm{v}$ is itself a JVP through the residual, requiring no explicit Jacobian assembly. The linear solve against $\bm{K}$ reuses the factorization from the final Newton step. Both IFT rules are registered as custom JVP primitives within JAX \cite{bradbury2018jax}, composing naturally with its reverse-mode AD engine.

\section*{Declaration of Competing Interest}

The authors declare that they have no known competing financial interests or personal relationships that could have appeared to influence the work reported in this paper

\section*{Compliance with ethical standards}

The authors declare that they have no conflict of interest.

\section*{Acknowledgments}

A.C. and W.C. acknowledge the support of DARPA METALS under agreement No. HR0011-24-2-0302. W.C., D.D.L., and H.K. acknowledge support from the NSF Boosting Research Ideas for Transformative and Equitable Advances in Engineering (BRITE) Fellow Program (CMMI-2227641). W.C. and  D.D.L. are also grateful for support from NASA's Minority University Research and Education Project Institutional Research Opportunity (MIRO) through the Center for In-Space Manufacturing (CISM-R2): Recycling and Regolith Processing (Award 80NSSC24M0176). W.C. acknowledges support from the NSF Materials Research Science and Engineering Centers (MRSEC) program (DMR-2308691). The authors also acknowledge Shicheng Li and Deepak Sharma for their inputs.

\section*{Declaration of generative AI and AI-assisted technologies}

During the preparation of this manuscript, the authors used Google’s Gemini and Anthropic’s Claude to improve the language and readability. After using these tools, the authors reviewed and edited the content as needed and take full responsibility for the content of the publication.

\section*{Replication of Results}
The Python source code is available at \href{https://github.com/ideal-nu/gelato}{github.com/ideal-nu/gelato}. The implementation is also provided as supplementary material to the version published in the journal.

\bibliographystyle{unsrt}  
\bibliography{7_references}

\end{document}